\documentclass[iop]{emulateapj}

\usepackage{natbib,graphicx}
\begin{document}
\title{The 2011 Outburst of Recurrent Nova T~Pyx:\\
 Radio Observations Reveal the Ejecta Mass and Hint at Complex Mass Loss}
\author{Thomas Nelson\altaffilmark{1}, Laura Chomiuk\altaffilmark{2,3}, Nirupam Roy\altaffilmark{3,7}, J.~L.~Sokoloski\altaffilmark{4}, Koji Mukai\altaffilmark{5,6}, Miriam I.~Krauss\altaffilmark{3}, Amy J.~Mioduszewski\altaffilmark{3},  Michael P.~Rupen\altaffilmark{3},  \& Jennifer Weston\altaffilmark{4}}
\altaffiltext{1}{School of Physics and Astronomy, University of Minnesota, 115 Church St SE, Minneapolis, MN 55455}
\altaffiltext{2}{Department of Physics and Astronomy, Michigan State University, East Lansing, MI 48824, USA}
\altaffiltext{3}{National Radio Astronomy Observatory, P.O. Box O, Socorro, NM 87801, USA}
\altaffiltext{4}{Columbia Astrophysics Laboratory, Columbia University, New York, NY, USA}
\altaffiltext{5}{Center for Space Science and Technology, University of Maryland Baltimore County, Baltimore, MD 21250, USA}
\altaffiltext{6}{CRESST and X-ray Astrophysics Laboratory, NASA/GSFC, Greenbelt MD 20771 USA}
\altaffiltext{7}{Now at Max Planck Institut f\"{u}r Radioastronomie, Auf dem H\"{u}gel 69 53121 Bonn, Germany}
\email{tnelson@physics.umn.edu}

\begin{abstract}
Despite being the prototype of its class, T~Pyx is arguably the most unusual and poorly understood recurrent nova.  Here, we use radio observations from the Karl G.~Jansky Very Large Array to trace the evolution of the ejecta over the course of the 2011 outburst of T~Pyx.  The radio emission is broadly consistent with thermal emission from the nova ejecta. However, the radio flux began rising surprisingly late in the outburst, indicating that the bulk of the radio-emitting material was either very cold, or expanding very slowly, for the first $\sim$50 days of the outburst.  Considering a plausible range of volume filling factors and geometries for the ejecta, we find that the high peak flux densities of the radio emission require a massive ejection of $(1-30) \times 10^{-5}$ M$_{\odot}$.  This ejecta mass is much higher than the values normally associated with recurrent novae, and is more consistent with a nova on a white dwarf well below the Chandrasekhar limit. \end{abstract}

\section{Introduction}
\label{intro}
Nova outbursts are the most common stellar explosions in the universe, and are important sources of both energy and matter injection into the interstellar medium (ISM; \citealt{Gehrz98}). Novae  are understood to be the result of a thermonuclear runaway on the surface of an accreting white dwarf. This runaway begins when the critical pressure and temperature required for CNO cycle burning of hydrogen are reached at the base of the accreted shell \citep{Schatzman49, Starrfield72}.  As long as accretion continues, all novae are expected to recur on timescales from $\sim$10$^8$ years to less than a year \citep[e.g.,][]{Yaron_etal05}.  This recurrence time depends primarily on the mass of the white dwarf and the accretion rate from the binary companion; more massive white dwarfs have smaller radii and therefore higher surface gravities, meaning that the critical conditions are reached for smaller accreted envelopes.  The accretion rate controls the time required to accumulate the critical mass---higher accretion rate systems accrete the trigger mass in less time.  

One of the most important predictions of theoretical studies of novae is that most white dwarfs experience a net loss in mass during outburst \citep{MacDonald84, Yaron_etal05}, implying severe difficulties in growing white dwarfs to the Chandrasekhar mass so that they explode as supernovae of Type Ia \citep[e.g.,][]{Livio01, Maoz14}.  Only white dwarfs accreting material at very high rates from their companions are predicted to eject less mass than had been accreted in the quiescence period prior to outburst \citep[e.g.,][]{Yaron_etal05}.  Such high accretion rates are also expected to result in relatively short intervals between outbursts, with massive white dwarfs experiencing outbursts every decade or so. These characteristics are observed in recurrent novae (i.e., the subclass of novae with more than one recorded outburst).  Recurrent novae present the best opportunity to determine whether---and in what circumstances---accreting white dwarfs can be Type Ia supernova progenitors, by measuring which is larger: the mass ejected in the nova outburst or the mass accreted since the last outburst.  However, the models of \citet{Yaron_etal05} find that the difference between the accreted and ejected mass is seldom greater than a factor of $\sim$2. Therefore to truly constrain whether a white dwarf is growing or shrinking in mass, we need accurate and precise measurements of both the accretion rate and the ejecta mass. Radio observations are key in determining the latter \citep{Seaquist_Bode08}.

\begin{figure*}
\hspace{0.5cm}
\includegraphics[width=7.5in]{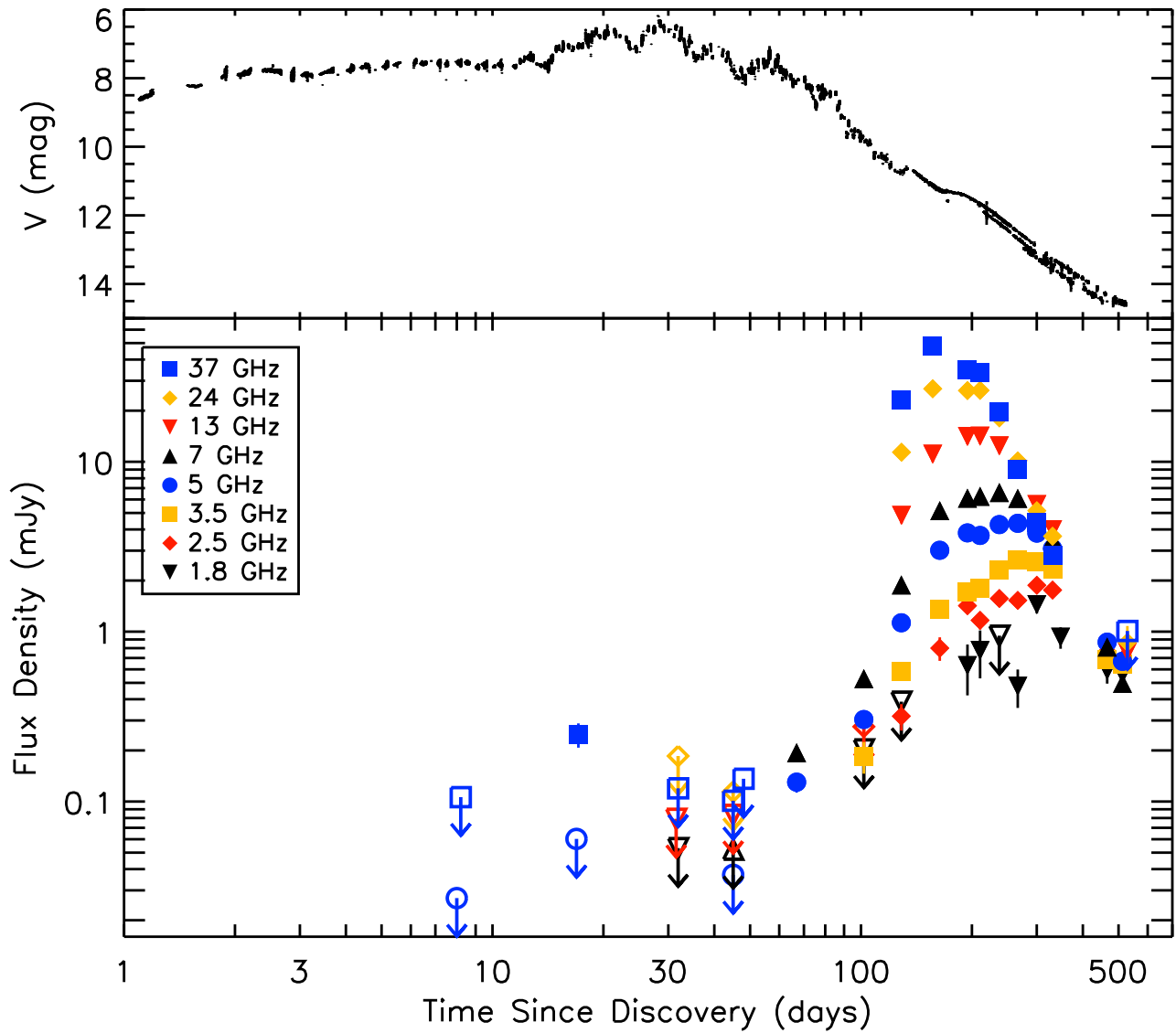}
\caption{Optical and radio light curves of the 2011 T~Pyx outburst.  Upper panel: V band optical light curve from AAVSO data.  Lower panel: VLA light curves at frequencies from 1.8 to 33 GHz. Open symbols with arrows indicate 3-$\sigma$ upper limits on the flux density.  We take 2011 April 14 as Day 0, the start of the outburst. }
\label{ro}
\end{figure*}

T~Pyx is the prototype recurrent nova, with outbursts observed in 1890, 1902, 1920, 1944, and 1966.  However, despite its canonical status and more than a century of study, many questions about this system remain unanswered.  T~Pyx differs from most other recurrent novae in a number of important ways.  The outburst evolution is much slower, taking longer to both rise to maximum light and subsequently fade back to quiescence \citep{Schaefer10b}. T~Pyx is the only recurrent nova with an orbital period below the period gap ($P = 1.83$ h; \citealt{Schaefer92, Patterson98, Uthas10}), with IM Nor coming a close second at 2.46 h (in the period gap).  Mass transfer in such short-period systems is driven by angular momentum loss due to gravitational radiation, which typically leads to accretion rates of order 10$^{-10}$ M$_{\odot}$ yr$^{-1}$ or less \citep{Knigge11}. However, there is strong evidence for a high accretion rate in T~Pyx \citep{Selvelli08, Godon14}.
Finally, the binary system is surrounded by a clumpy remnant  that is not observed in other recurrent systems, extending up to 10$^{\prime\prime}$ from the central binary \citep{Williams82, Shara_etal97, Schaefer10a}. 

These unusual characteristics have led to questions about the evolutionary state of T~Pyx.  The short orbital period and low mass of the companion star make it difficult to maintain the high accretion rate required to power a nova outburst every few decades.  To solve this problem, authors have suggested that a normal nova outburst in the late 1800s led to a period of enhanced mass loss induced by irradiation of the companion star by the hot, shell burning white dwarf \citep[e.g.,][]{Knigge00, Schaefer10a}.  This higher mass loss rate in turn fed a higher accretion rate onto the white dwarf, and led to the accumulation of the critical mass required for an outburst in just a decade or so.  There is evidence that the average accretion rate between outbursts has been declining over the 20th century, and that the recurrence time between novae has been growing, suggesting that the current state of the T~Pyx system is unsustainable for extended periods of time \citep{Schaefer10a}.  

T~Pyx experienced its most recent outburst in April 2011---the first since 1966, and thus the first eruption for which the astronomical community has been able to obtain broad multiwavelength coverage including the radio regime.  The $V$-band optical light curve of the outburst, created using data from the American Association of Variable Star Observers (AAVSO), is shown in the upper panel of Figure \ref{ro}.  The optical evolution of T~Pyx during outburst is much slower and more complex than that observed in most other recurrent novae.  The system brightened quickly from its quiescent value of $V \approx 15.5$ mag to $V \approx 8$ mag within the first 2 days of outburst. This initial brightening was then followed by a phase of remarkably constant flux, during which the optical brightness remained at $\sim$8 mag until Day $\sim$15.  T~Pyx then began to show variations around this level, reaching a maximum brightness of $V \approx 6.5$ mag around Day 30 but fluctuating around a flux plateau until Day $\sim$90 at which time the optical light curve suddenly declined. At the time of our last radio observations (late September 2012, roughly 1.5 years after the outburst) the V band magnitude as reported by AAVSO observers was $\sim$14.5 mag, indicating that the system has not yet fully returned to its quiescence flux levels \citep{Schaefer_etal11}.

The radio emission from most novae is thermal emission from the ionized expanding ejecta expelled in the nova event itself \citep{Seaquist77,Hjellming79,Seaquist80}, although a few novae with unusually dense environments have shown synchrotron emission or thermal emission from the circumbinary medium \citep{Kantharia07, Rupen08, Chomiuk_etal12}. In the classical picture of thermal emission from novae, a multi-frequency radio light curve measures the entire density profile of the ejecta by monitoring the recession of the radio photosphere. Radio observations therefore trace the bulk of the ejected material in a relatively straightforward way, enabling us to derive the total mass of a nova's ejecta.  For a detailed discussion of radio emission from novae, see \citet{Seaquist_Bode08}.

In this paper we present multifrequency radio light curves of the 2011 nova outburst in T~Pyx, obtained with the newly-upgraded Karl G.~Jansky Very Large Array (VLA) of the National Radio Astronomy Observatory, and use them to determine the mass of the ejected material. We present our observations and data analysis in Section 2.  Section 3 describes the evolution of the radio emission over the course of the outburst, and investigates the implications of the optically-thick rise of the radio light curve.  In Section 4, we estimate the mass of the ejecta and the timing of the ejection. Section 5 explores and rules out other possible scenarios which might explain the radio emission from T~Pyx and in Section 6 we discuss the implications of the ejecta  mass measurement . Finally, we present our conclusions in Section 7.  

Throughout this paper, we assume a distance to the system of 4.8 $\pm$ 0.5 kpc, derived from light echo observations with the {\it Hubble Space Telescope (HST)} during the 2011 outburst \citep{Sokoloski13}, and consistent with the distance of $>$4.5 kpc inferred from optical spectroscopy by \citet{Shore11}. We take the first day of dramatic optical brightening, 2011 April 14.0 UT or 55665.0 MJD \citep{Waagan_etal11}, as the start of the outburst and label it as Day 0.

\begin{figure}
\vspace{-0.6cm}
\begin{center}
\includegraphics[width=3.7in, angle=90]{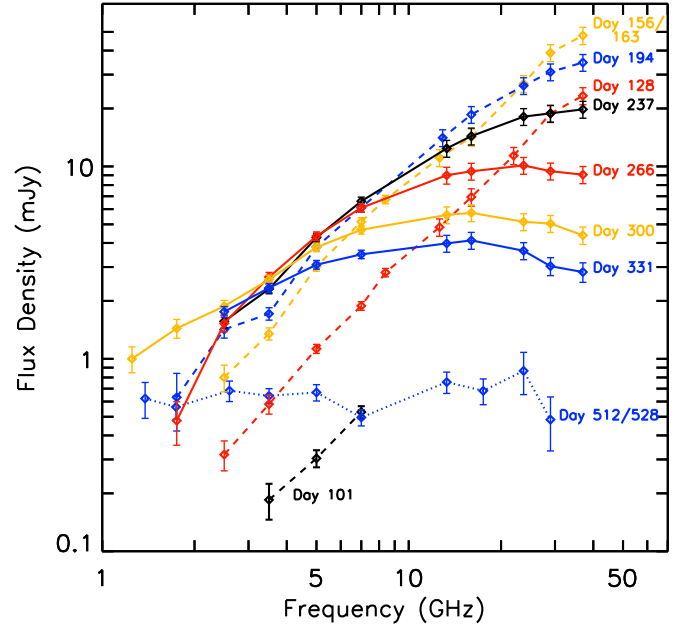}
\caption{Time evolution of the radio spectrum of T~Pyx, spanning Days 101--512.}
\label{forbob}
\end{center}
\end{figure}

\section{Observations and Data Reduction}
\label{data}

We obtained sensitive radio observations of T~Pyx between 2011 April 22 and 2012 September 23 with the VLA through programs 11A-263, 11B-230, and 12A-459.  Over the course of the outburst, the VLA was operated in all configurations, and data were obtained in the L (1--2 GHz), S (2--4 GHz), C (4--8 GHz), X (8--8.8 GHz), Ku (12--18 GHz), K (18--26.5 GHz), and Ka (26.5--40 GHz) bands, resulting in coverage from 1--37 GHz.  Observations were acquired with 2 GHz of bandwidth (the maximum available), split between two independently tunable 1-GHz-wide basebands, with the exception of L-band observations, which provide 1 GHz of bandwidth between 1--2 GHz, and X-band observations, which were still subject to the old receivers and therefore only cover 800 MHz of bandwidth.  The details of our observations are given in Table 1.

\begin{figure*}
\begin{center}
\includegraphics[width=5.0in,angle=90]{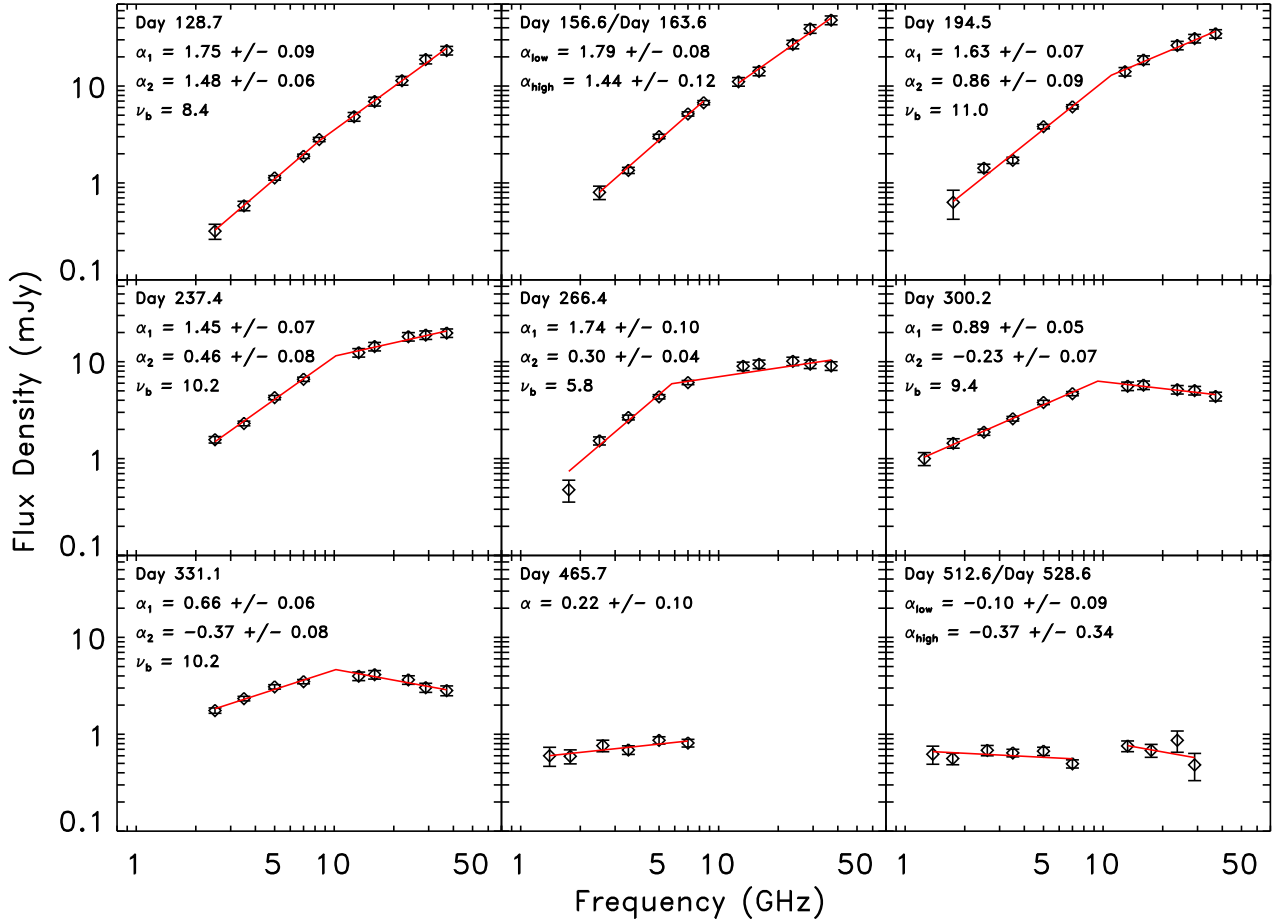}
\caption{A selection of VLA radio spectra over the evolution of the outburst.  At each epoch, we over-plot the best-fit single or broken power law spectrum in red.  A single index quoted in the top-left corner indicates that a single power law is sufficient to fit the data. Where two indices are quoted, $\alpha_1$ describes the lower frequency data while $\alpha_2$ is the fit to the higher frequency data; the break frequency (in GHz) is denoted as $\nu_b$. Where low- and high-frequency data are obtained at significantly different times, we fit them separately and report two indices.}
\label{blfit}
\end{center}
\end{figure*}

At the lower frequencies (L, S, and C bands), the source J0921-2618 was used as the gain calibrator, while J0900-2808 was used for gain calibration at the higher frequencies (X, Ku, K, and Ka bands). The absolute flux density scale and bandpass were calibrated during each run with either 3C138, 3C147, or 3C286. Referenced pointing scans were used at Ku, K, and Ka bands to ensure accurate pointing; pointing solutions were obtained on both the flux calibrator and gain calibrator, and the pointing solution from the gain calibrator was subsequently applied to our observations of T~Pyx. Fast switching was used for high-frequency calibration, with a cycle time of $\sim$2 minutes.  Data reduction was carried out using standard routines in AIPS and CASA. Each receiver band was edited and calibrated independently. The calibrated data were split into their two basebands and imaged, thereby providing two frequency points, with the exception of X band (where only 800 MHz of bandwidth is obtained, centered at 8.4 GHz) and K band (where the two basebands were placed side-by-side and averaged together). In most cases, when there was sufficient signal, a single iteration of phase-only self-calibration was carried out, with a solution interval of 1--5 minutes, using the clean components from the first imaging run.

One complication affecting our data was that A configuration (the most extended VLA configuration) occurred during the summer months (June--August 2011), when the VLA site is subject to poor weather. These adverse conditions, combined with the low elevation of T~Pyx, led to severe decorrelation of many of our A configuration observations, particularly at high frequencies. The K and Ka band data, obtained on 2011 August 20 and September 17, were in particular very badly decorrelated.  Despite this problem, significant flux density was clearly detected; a point source model was used in self-calibration to recover the flux. 

In each image, the flux density of T~Pyx was measured by fitting a gaussian to the imaged source with the tasks \verb|JMFIT| in AIPS and \verb|gaussfit| in CASA. We record the integrated flux density of the gaussian; in most cases, there was sufficient signal on T~Pyx to allow the width of the gaussian to vary slightly, but in cases of low signal-to-noise ratio, the width of the gaussian was kept fixed at the dimensions of the synthesized beam. Errors were estimated by the gaussian fitter, and added in quadrature with estimated calibration errors of 5\% at lower frequencies ($<$10 GHz) and 10\% at higher frequencies ($>$10 GHz).  All resulting flux densities and uncertainties are presented in Table \ref{tab:phot}.              

\section{Radio evolution of the 2011 Outburst of T~Pyx}
\begin{figure}
\vspace{-0.3cm}
\hspace{-0.4cm}
\includegraphics[width=3.7in,angle=90]{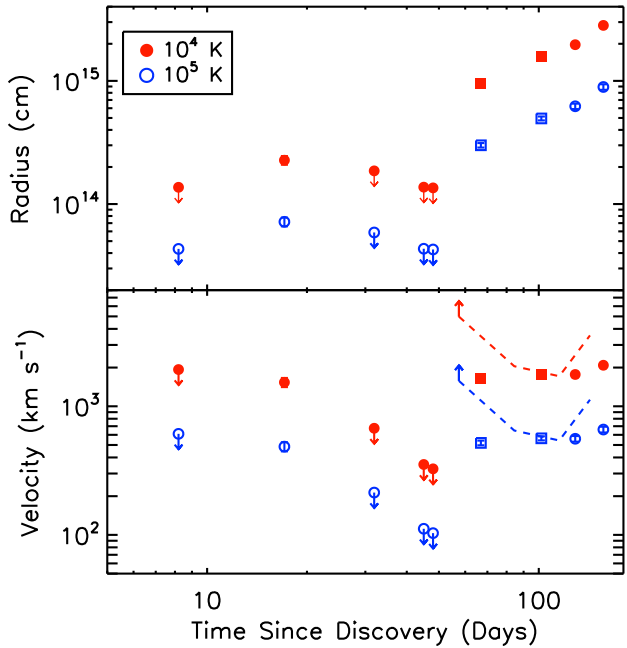}
\caption{{\bf Top Panel:} Radius of the nova ejecta as a function of time from outburst, assuming that the radio emission is optically thick, thermal, and of constant temperature. Filled red points assume an electron temperature of $10^{4}$ K, while open blue points represent $10^{5}$ K. For these values of ejecta temperature, we find that the ejecta must be confined to a small radius ($\lesssim10^{14}$ cm) on Day 48 in order to be consistent with radio non-detections. Subsequently, the implied ejecta radius rapidly increases as the radio light curve rises.
{\bf Bottom Panel:} Expansion velocity of the nova ejecta, calculated using the radii from the top panel. Filled red and open blue points are for ejecta temperatures of $10^4$ and $10^5$ K respectively, and assume that the ejecta begin expanding on Day 0 (2011 April 14). Dashed lines represent instantaneous velocities calculated between two adjacent radio detections. 
The 33 GHz measurements (circles) are most constraining when they are available, but they were not obtained on Days 66 and 102, so we use 7 GHz (squares) on these dates.}
\label{thickrad}
\end{figure}

The VLA light curves and spectra, presented in Figures \ref{ro}, \ref{forbob}, and \ref{blfit}, show the evolution of the radio emission over the course of the outburst.  The source was not detected at either 33 or 6 GHz during the first observation, on Day 8, with upper limits on the flux density $\leq$0.1 mJy (Table \ref{tab:phot}).  Subsequently, on Day 17, T~Pyx was detected at 33 GHz at a significance of 8$\sigma$, but was not detected at 6 GHz. In the next two epochs (Days 31 and 45/48), we observed with additional receiver bands, but the source was not detected at any frequency.  Assuming an optically thick spectrum, the most constraining flux density upper limits at this time were obtained at high frequencies, again with $F_{\nu}$ $\leq$ 0.1 mJy.  On Day 66, observations at 5 and 7 GHz finally resulted in detections of the source at the $\sim$0.1 mJy level, and in subsequent observations T~Pyx brightened at all frequencies, reaching a maximum flux density of 48 mJy at 37 GHz on Day 156.  The radio flux began to fade after this date, first at the highest frequencies, and later at lower frequencies.  By Day 528 (the final epoch of observations) the flux density had declined at all frequencies to a level of 0.5--0.9 mJy.

\subsection{Spectral Evolution}
We modeled the spectrum obtained at each epoch as a single power law of the form $S_{\nu} \propto \nu^{\alpha}$ (where $\alpha$ is the spectral index). When flux density measurements were obtained at more than five frequencies, we also fit a broken power law.  The fitting procedure was carried out using {\tt mpfit}, a curve fitting package implemented in IDL \citep{Markwardt09}. For each spectrum, we compared the goodness of fit of the single and broken power law models, as measured by $\chi^{2}$/$\nu$ (where $\nu$ is the number of degrees of freedom in the model). We selected the model with the lower value of reduced $\chi^{2}$ as the best fit, and plotted it as a red line in each panel of Figure~\ref{blfit}. During the early epochs, the broken power law model provides only a marginally better fit to the data than the single power law.  However at later times, where the turnover at higher frequencies is clear, the broken power law provides a large improvement in the goodness of the fit.  The fit on Day 266 is the worst; on this day, the spectrum is clearly curved at high frequencies, and additional power laws would be required to obtain a statistically good fit.

The radio spectra indicate an evolution from a source suffering strong free-free absorption (during the rising phase of the light curve), to an optically-thin thermally emitting source at late times.  We present a selection of spectra that best demonstrates this evolution in Figures \ref{forbob} and \ref{blfit}. The spectral indices fit to both the low- and high-frequency data are initially steeply rising ($\alpha_{1}$ =1.75 $\pm$ 0.09,  $\alpha_{2}$ =1.48 $\pm$ 0.06 on Day 128), indicative of optically thick emission (for which $\alpha$ = 2.0 is theoretically expected, while slightly lower spectral indices---$\alpha \approx 1.5$---are commonly observed in novae at early times; T.~Johnson et al.\ 2014, in preparation).  We note that the spectral index implied by the data on Day 17 ($\alpha$ $\geq$ 1.14) is also steep, and suggests that the source was also optically thick at the time of first detection.  

As the outburst evolves the spectrum becomes shallower, beginning first at high frequencies on Day 194, and eventually at low frequencies by Day 331 (Figure \ref{forbob}). The best fit indices at high frequencies gradually decline from $\alpha$ = 0.9 (Day 194) to 0.3 (Day 266), consistent with expectations for a partially optically thick shell with a density profile that decreases with radius.  By Day 300, the spectral index at high frequencies is slightly negative ($\alpha = -0.23 \pm 0.07$), consistent with emission from a completely optically thin shell (theoretically expected to be $\alpha = -0.1$).  By the epochs of our last observations (Days 512 and 528), the spectra have indices expected for optically thin gas at all frequencies.  

\subsection{Evolution While T~Pyx is Optically Thick at Radio Wavelengths}\label{optthick}

The steep radio spectra measured during the rising part of the radio light curve are consistent with optically-thick thermal emission, and therefore the radio flux density is a simple function of the angular size and temperature of the ejecta at these early times.  For an optically-thick thermal source, the flux density can be expressed as:
\begin{equation}
S_{\nu} = {{T\, \theta^2} \over {{1.96\, \lambda^{2}}}}\ {\rm mJy}
\end{equation}
where $\lambda$ is the observing wavelength in cm, $T$ is the temperature of the radio photosphere in K, and $\theta$ is the angular diameter of the radio photosphere in arcsec.

We can extract information about the properties of the radio emitting region in T~Pyx from the early radio light curve using two diametrically-opposed assumptions: either holding temperature fixed, and calculating variations in the radius of the photosphere, or holding expansion velocity fixed and measuring the photospheric temperature. First, we hold temperature fixed, as plotted in Figure \ref{thickrad}. Brightness temperatures derived from spatially-resolved radio images of novae are in the range 10$^{4}$--10$^{5}$ K \citep{Taylor88, Pavelin93, Eyres96, Heywood05}, consistent with a photoionized gas. Assuming a distance of 4.8 kpc, we obtain estimates of the photospheric radius for temperatures in this range (the difference between a $10^{4}$ K and a $10^{5}$ K photospheric radius is a constant factor of $\sqrt{10} \approx 3.2$).
We calculate upper limits on radius for epochs where we only measure upper limits on flux density (Figure \ref{thickrad}, top panel). 

We find that if the ejecta are warm, they must have a very small radius on Day 48 ($\lesssim10^{14}$ cm) in order to be consistent with the radio non-detections observed on this date. Assuming these ejecta were launched from the white dwarf on Day 0, the inferred expansion velocities are very low ($\lesssim$ few hundred km s$^{-1}$, see Figure \ref{thickrad}, bottom panel). Starting with the radio detection on Day 66, the derived radius suddenly increases by an order of magnitude (in just 18 days), implying an epoch of fast expansion with instantaneous velocities $>$2,000 km s$^{-1}$ (dashed lines in the bottom panel of Figure \ref{thickrad}).  In subsequent observations, the radii increase at a slower rate, giving expansion velocities of $\sim$1,000 km s$^{-1}$ between days 67 and 128.  Between the last two optically thick epochs (Days 128 and 156/174), the inferred expansion velocity increases again, by about a factor of 2 compared to the previous pair of measurements. Based on this evolution, we conclude that the radio light curve is completely inconsistent with free expansion at a single velocity from Day 0 if we require that the ejecta temperature remains warm and relatively constant.

\begin{figure}[t]
\begin{center}
\includegraphics[height=3.4in,angle=90]{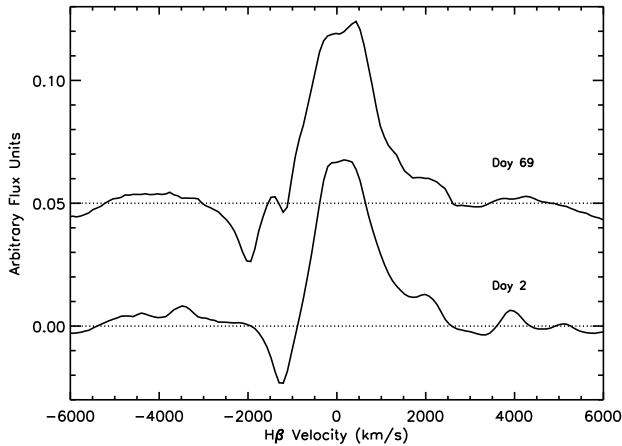}
\caption{H$\beta$ line profiles for T~Pyx observed 2 and 69 days after the beginning of optical rise. The P~Cygni absorption becomes more blue-shifted with time, increasing from 1,900 km s$^{-1}$ on Day 2 to 3,000 km s$^{-1}$ on Day 69. Spectra kindly provided by Stonybrook/SMARTS Atlas of Southern Novae \citep{Walter12}.}
\label{hbeta}
\end{center}
\end{figure}

Examining the H$\beta$ line profile on Day 2 (Figure \ref{hbeta}; \citealt{Walter12}), we estimate that the blue wing of the P~Cygni profile yields a maximum expansion velocity of $\sim$1,900 km s$^{-1}$. On Day 69 (around the time that the radio light curve starts to rise), the H$\beta$ absorption component has become more blue-shifted, now implying an expansion velocity of $\sim$3,000 km s$^{-1}$. A similar increase in the expansion velocity of T~Pyx are measured by \citet{Surina14} in the first two months of outburst. The velocities derived from the H$\beta$ line profiles, and those implied by the measured flux densities, are therefore not compatible---the optical spectra predict much larger physical sizes of the ejecta around Day 50, 
and therefore higher flux densities, if the ejecta temperature is in the range 10$^{4}-10^{5}$ K.  Could changing temperature in the ejecta therefore account for the unusual flux density evolution?

\begin{figure}
\vspace{-0.3cm}
\hspace{-0.4cm}
\includegraphics[width=2.7in,angle=90]{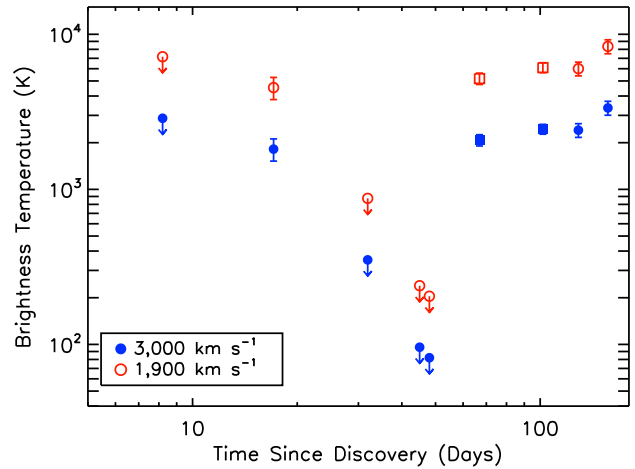}
\caption{Brightness temperature (or upper limits) as a function of time from outburst, assuming T~Pyx is emitting optically thick thermal radio emission during the first 180 days. We assume constant expansion velocity of either 3,000 km s$^{-1}$ (filled blue points) or 1,900 km s$^{-1}$ (open red points), starting on Day 0 (2011 April 14). If the bulk of the mass was ejected on Day 0, it must cool dramatically to be consistent with our radio upper limits, reaching $\sim$100 K on Day 48, and then increase suddenly to a few thousand K. In most epochs, the 33 GHz measurements are most constraining on the temperature (plotted as circles), but on Days 66 and 102 we lack high-frequency observations, and therefore plot 7 GHz measurements instead, as squares.}
\label{thicktemp}
\end{figure}

If we assume that we know the physical size of the ejecta (given simply by $v t$), we can estimate the temperature of the ejecta.  In Figure \ref{thicktemp}, we show the derived brightness temperatures assuming expansion of the ejecta from Day 0 of 1,900 km s$^{-1}$ or 3,000 km s$^{-1}$ (the range of velocities found from the H$\beta$ lines above; yielding temperatures that span a factor of $(3,000/1,900)^2 \approx 2.5$).  Under these assumed velocities, the flux density limits imply a dramatic cooling of the ejecta during the first two months of outburst. The derived temperatures are as low as 100 K on Day 48 assuming an expansion velocity of 3,000 km s$^{-1}$.   This cooling is followed by a dramatic re-heating of the ejecta by Day 67, when the inferred temperature has increased to a few thousand K. 

To summarize, the deep radio flux limits at early times, and the observed sudden brightening around Day 67, are not consistent with the free expansion of a warm shell of gas from Day 0.  If the shell is warm, there must be a period of dramatic increase in the expansion velocity.  Alternatively, if the ejecta expand at constant velocity from the onset of the outburst, a period of dramatic cooling and re-heating is required to account for the radio light curve.  The radio data alone cannot distinguish between the two possibilities, although insights from X-ray and optical observations can break this degeneracy (see \citealt{Chomiuk13} for more details).

\section{Ejecta mass and ejection timing} \label{model}
Despite the unusual rate of the radio flux density rise, the evolution of the radio flux and spectra are generally consistent with the standard picture of thermal emission from expanding nova ejecta. The early-time emission is consistent with expectations of a compact, optically-thick shell.  At this stage, the radio photosphere (i.e. the surface where optical depth $\tau_{\nu}$ = 1) is essentially coincident with the outer surface of the ejecta for all radio frequencies, and expansion of the ejecta leads to increasing flux.  

\begin{figure}[t]
\begin{center}
\includegraphics[width=3.5in, angle=90]{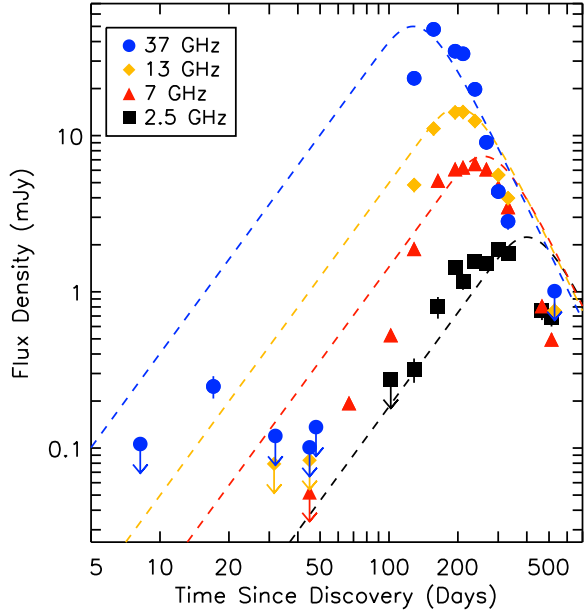} \\
\caption{Homologous expansion model (shown as dashed lines, one per observed frequency) for the radio emission from the nova ejecta, assuming mass ejection at the time of optical discovery.  The data are shown as solid points and lines.  We assume a distance to T~Pyx of 4.8 kpc, and expansion velocity of 3,000 km s$^{-1}$.  The model plotted here has ejected mass of $3 \times 10^{-4}\ M_{\odot}$, $T_{e} = 10,000$ K, and $v_{\rm min}/v_{\rm max} = 0.9$.  The match to the high flux densities observed at peak brightness is relatively good, but the model fails to reproduce all other features of the radio evolution.}
\label{lcmodel_t0}
\end{center}
\end{figure}

The subsequent flattening of the radio spectrum observed at later times is also compatible with an expanding thermal shell.  As the density in the shell drops, so does the free-free optical depth:
\begin{equation}
\tau_{\nu} = 8.235 \times 10^{-2} \left({{T_e}\over{\rm K}}\right)^{-1.35} \left({{\nu}\over{\rm GHz}}\right)^{-2.1} \left({{\rm EM}\over{\rm cm^{-6}\ pc}}\right)
\end{equation}
\citep{Lang80}. Here, EM is the emission measure, defined as $\int n_{e}^2\,dl$, where $n_{e}$ is the electron density of the absorbing material and $dl$ is the path length through this material. The electron temperature is denoted as $T_e$, and the frequency of observations is $\nu$. The transition to optically thin emission begins at high frequencies (since higher radio frequencies require larger emission measures to be optically thick compared to lower frequencies), and then cascades to lower frequencies.

\begin{figure}[t]
\begin{center}
\includegraphics[width=3.5in, angle=90]{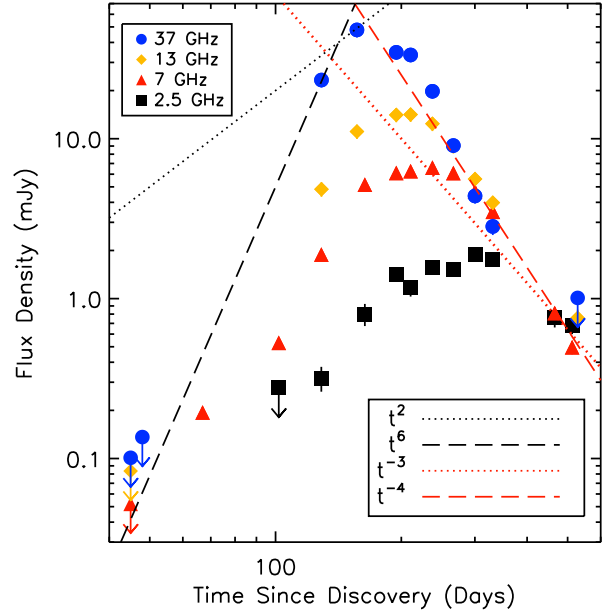} \\
\caption{Thermal radio emission from an isothermal shell expanding since Day 0 should yield a power-law rise in the radio light curve, $S_{\nu} \propto t^2$ (black dotted line) and a power-law decline, $S_{\nu} \propto t^{-3}$ (re dotted line). Instead, the radio light curve of T~Pyx is better described as $S_{\nu} \propto t^6$ (black dashed line) on the rise and $S_{\nu} \propto t^{-4}$ (red dashed line) on the decline.}
\label{lcpl}
\end{center}
\end{figure}

The radio photosphere at a given frequency will recede though the ejecta, assuming the ejecta are distributed in a thick shell.  This process leads to a gradual turnover in the radio light curve, and a flattening of the spectral index (to $\alpha \approx 0.6$ for an $r^{-2}$ density profile; e.g., \citealt{Panagia_Felli75}). The spectral turnover cascades towards lower frequencies as time passes, just as we see in the data.   Once the photospheres of all observed frequencies have receded through the entire ejected shell, the emission can be described completely by optically-thin thermal bremsstrahlung ($\alpha = -0.1$). Subsequently, the radio light curve simply fades at all frequencies as the ejecta expand and the emission measure continues to drop.  This picture provides a good match to the late time spectral index and flux density evolution of T~Pyx.


We can estimate the mass of the ejecta using simple models that account for the observed radio flux evolution at all frequencies in a consistent manner.  As a starting point, we follow the work of \citet{Seaquist77} and \citet{Hjellming79} and assume that the ejecta can be described as a thick, spherically symmetric shell of warm (T $\approx$ 10$^{4}$ K) material emitting thermal bremsstrahlung radiation (we will discuss the effects of non-spherical geometries, undoubtedly present in T Pyx, in Section 4.3).  The density profile of the spherical shell decreases with radial distance such that $\rho(r) \propto r^{-2}$ or $r^{-3}$, with an inner boundary to the ejecta.  The thickness of the ejected shell may be due to either a velocity spread in a homologous explosion or a prolonged period of mass loss \citep{Kwok83}.  

The model traces the evolution of the photosphere at each frequency as the ejecta expand over time, and depends on the ejected mass ($M_{\rm ej}$), the velocity of the outermost ejecta ($v_{\rm max}$), the electron temperature of the ejecta ($T_{e}$; assumed to be constant in time and throughout the ejecta), and the distance to the nova ($d$).  The turn-over in the light curves also depends on the assumed density profile (we take $\rho(r) \propto r^{-2}$ here) and on the thickness of the shell, which we parameterize here as the ratio of the velocities in the innermost and outermost ejecta ($v_{\rm min}/v_{\rm max}$), assuming an instantaneous and homologous explosion.  We fit this model of expanding thermal ejecta to all observed frequencies simultaneously with the singular exclusion of the lower baseband in L band ($\sim$1.3 GHz; this frequency mostly yields non-detections), making use of  the {\tt mpfit} package developed for IDL \citep{Markwardt09} to carry out least-squares fitting and obtain the best-fit parameters. Measurements that are not significant at the 3$\sigma$ level are excluded from our model fits.
  
\
No good fits to the data are found for ejecta models that start expanding on Day 0.  The poor fits are not surprising, given the discontinuous evolution of the flux densities before Day 70 (Section \ref{optthick}). In this simple model, the expansion velocity of the ejecta is constant with time, resulting in flux densities that increase as $t^{2}$, completely inconsistent with the observations (Figures \ref{lcmodel_t0} and \ref{lcpl}). In fact, the rise of the light curve is much better described as a $S_{\nu} \propto t^{6}$ power law (relative to Day 0), as shown in Figure \ref{lcpl}. Discrepancies between the model and observations are also apparent during the light curve decline. A shell expanding from Day 0 should adhere to $S_{\nu} \propto t^{-3}$ once it becomes optically thin \citep{Seaquist_Bode08}, but the light curve of T~Pyx follow a steeper power-law decline, closer to $S_{\nu} \propto t^{-4}$ (Figure \ref{lcpl}).

 However, the simple  model still provides insight into some properties of the ejecta despite the poor fit to the early part of the light curve.  Assuming a distance to T~Pyx of 4.8 $\pm$ 0.5 kpc \citep{Sokoloski13}, an electron temperature of 10,000 K, and a maximum expansion velocity of 3,000 km s$^{-1}$, we find that only models with $M_{\rm ej} \approx  3 \times 10^{-4}$ M$_{\odot}$ can reproduce the peak flux densities (Figure \ref{lcmodel_t0}).  Irrespective of the quality of the fits at early and late times, the requirement that the ejecta must be physically large and optically thick to produce the high observed radio flux densities at maximum indicates that the mass of the ejecta must be high.

\begin{figure}[t]
\begin{center}
\includegraphics[width=3.5in, angle=90]{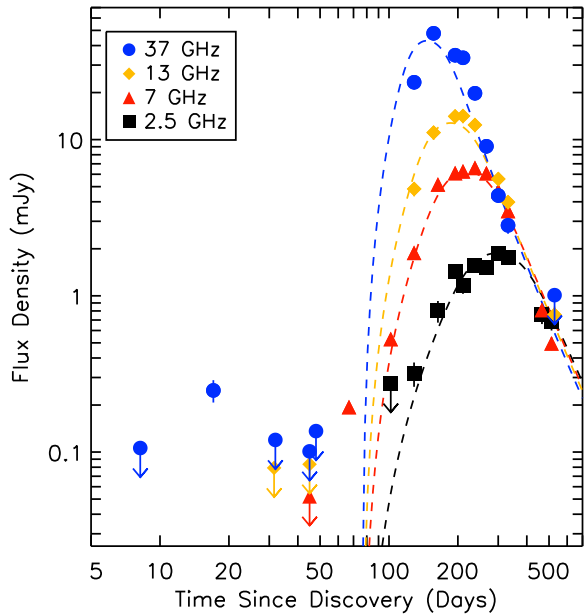} \\
\caption{Same as Figure \ref{lcmodel_t0}, but allowing the time of mass ejection to be an additional free parameter.  The formal best fit model has $t_{ej} = 76$ days, $M_{ej} = 2 \times 10^{-4}\ M_{\odot}$, and $T_{e} = 43,000$ K, and $v_{\rm min}/v_{\rm max} = 0.25$, although we note that the uncertainties on the ejection date are large.}
\label{lcmodel}
\end{center}
\end{figure}

\subsection{Accounting for low expansion velocities and/or low temperatures at early times} \label{delay}
One way to interpret the late, steep rise of the optically thick radio emission is with low expansion velocities before day 48, perhaps due to a delay in the ejection of the accreted shell.  We can incorporate this possibility into our simple shell model by including the time of ejection (measured relative to the date of discovery, 2011 April 14) as an additional free parameter.  The resulting model fits to the data are much better if the mass is ejected $\sim$75 days after the time of the optical discovery.  The best fit model for our assumed $v_{\rm max}$ and $d$  (3,000 km s$^{-1}$ and 4.8 kpc, respectively) is shown in Figure \ref{lcmodel}.  The ejected mass is $M_{\rm ej}$ = 2 $\times$ 10$^{-4}$, the ejection time $t_{0}$ is 76 days, the electron temperature of the ejecta is $T_{e} = 43,000$ K, and the shell has a thickness of $v_{\rm min}/v_{\rm max}=0.25$.  This model is consistent with the non-detections at early times, and provides a better fit to data at late times than the model in Figure \ref{lcmodel_t0}.  The relatively high electron temperature found by the model fit is required to account for both the high peak flux densities and the observed rate of transition to an optically-thin spectrum.

\begin{figure*}[t]
\begin{center}
\includegraphics[width=3.5in, angle=90]{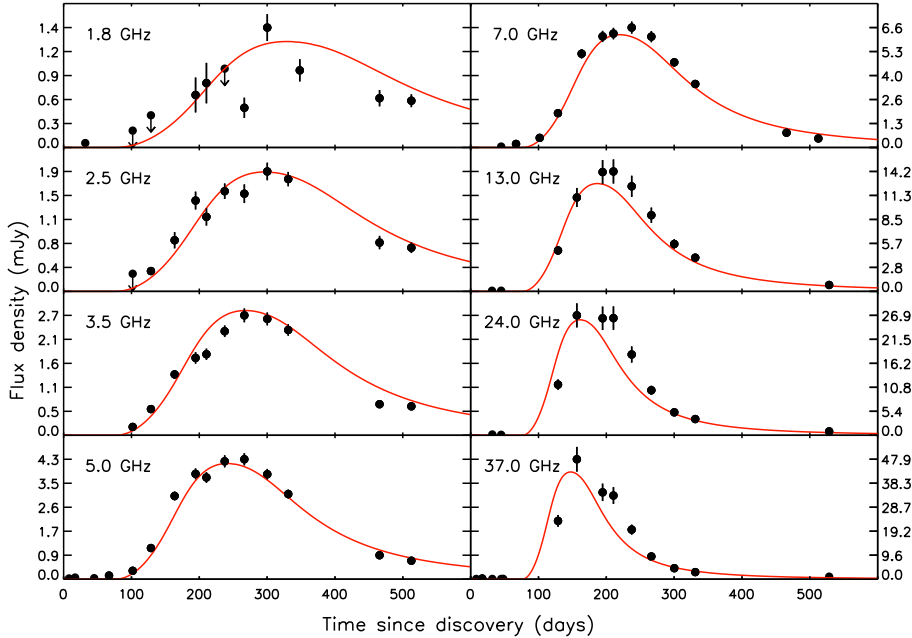} \\
\caption{VLA light curves of T~Pyx, with each frequency occupying its own panel. Red lines represent the model plotted in Figure \ref{lcmodel} and discussed in Section \ref{delay}. The error bars show the 1$\sigma$ uncertainty on the flux density.}
\label{lcbyfq}
\end{center}
\end{figure*}

The fit to the data is shown in more detail in Figure \ref{lcbyfq}, which plots the light curve at each frequency in separate panels. The model described above is over-plotted as a red line in each panel; in general the fit is good, especially considering the simplicity of our model. There is a significant detection of flux at 7 GHz on Day 67 that is significantly brighter than the model. The model begins to decline earlier at high frequencies than we observe in the data, although within 3$\sigma$ the points and the data agree. Some of this discrepancy could be due to effects of non-spherical geometry (see Section 4.3). Differences between the model and data could also be due to the assumption of a single temperature for the ejecta at all times and radii. 

The alternative interpretation of the low radio flux before Day 50 is that the ejecta are cold, $\sim$100 K.  The flux density of an optically-thick thermal body on the Rayleigh-Jeans tail varies linearly with the temperature.  If there is significant evolution in the temperature of the ejecta over the course of the outburst, this effect could account for a host of discrepancies between the model and the data.  Without additional constraints on either the temperature or the size of the ejecta, we cannot disentangle the trade-off between changes in these two parameters on the radio light curve. 
 
\subsection{Allowed ranges of the ejecta mass and electron temperature}
\begin{figure*}
\begin{center}
\includegraphics[width=4.5in, angle=90]{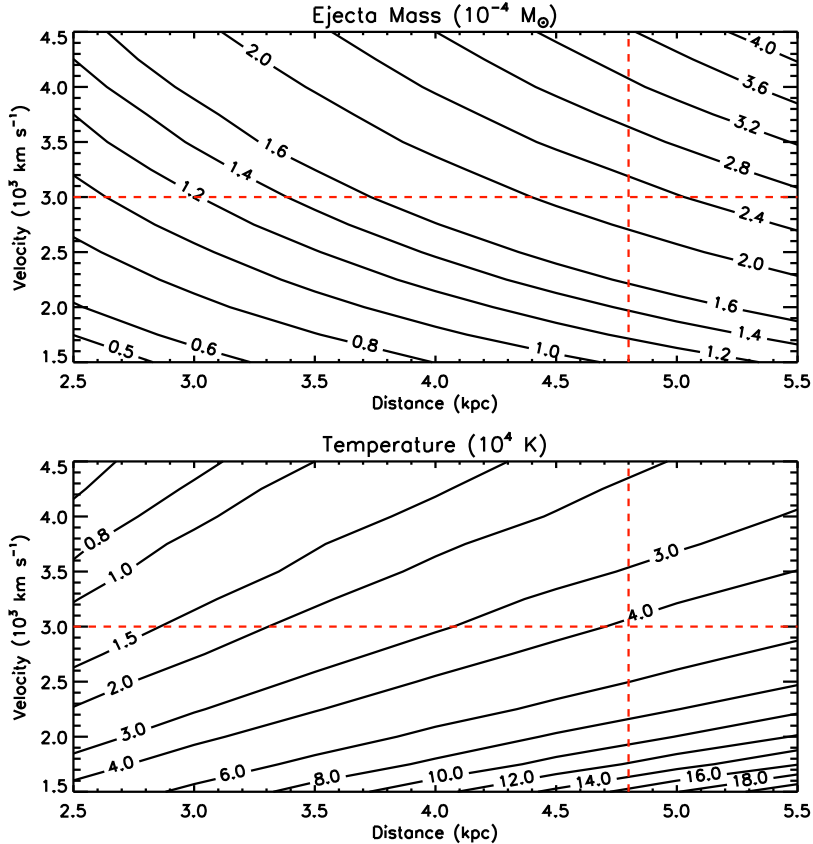}
\caption{Illustration of the degeneracies in our model fit to the radio light curves of T~Pyx. The derived values of ejecta mass ($M_{\rm ej}$; top panel) and electron temperature ($T_{e}$; lower panel) depend on assumed values for the distance to T~Pyx and the maximum expansion velocity in the nova outburst. Throughout this paper, we assume $d=4.8$ kpc and $v_{\rm max}$ = 3,000 km s$^{-1}$, as constrained by optical observations; these parameters are marked as dashed red lines. }
\label{fitcon_vd}
\end{center}
\end{figure*}

There are significant degeneracies between the four fundamental parameters of the model ($M_{\rm ej}$, $v_{\rm max}$, $T_{e}$; and $d$), as illustrated in Figure \ref{fitcon_vd}. Some of these parameters are poorly constrained by the data, for example $v_{\rm max}$, the maximum velocity of the ejecta. Although we take the value 3,000 km s$^{-1}$ (reported by \citealt{Osborne11} and estimated from H$\beta$ profiles on Day 69) as our fiducial value, a range of velocities have been reported in T~Pyx during the 2011 outburst \citep[e.g.,][]{Surina14}.  Similarly, although we favor the distance of 4.8 $\pm$ 0.5 kpc found using light echos from the nebular remnant observed with \emph{HST} (Sokoloski et al. 2012), distance estimates in the literature range from 2.5 kpc to $>$5 kpc. In order to understand the range of allowed ejected masses associated with these uncertainties, we explored a plausible parameter space of ejecta velocity and distance; for each combination we obtained the best-fit ejected mass and electron temperature, fitting to all frequencies simultaneously. Fits of similar quality (as defined by $\chi^2$ minimization) are found in all cases. The results are shown in Figure \ref{fitcon_vd}.

A change in the distance simply moves the model light curves up and down in brightness, but it influences the derived $M_{\rm ej}$, because more massive ejecta will have higher peak flux densities. Therefore, if T~Pyx is located at a smaller distance, we find a smaller ejecta mass from our radio light curves (top panel of Figure \ref{fitcon_vd}). However, a smaller ejecta mass will also become optically thin more quickly because the density in the shell is lower. The electron temperature varies along with ejecta mass because higher temperatures make the ejecta brighter while they are optically thick, but also lead to them becoming optically thin faster. Therefore, $M_{\rm ej}$ and $T_{e}$ are somewhat inversely related in the model fits, as seen by comparing the two panels in Figure \ref{fitcon_vd}. Finally, the expansion velocity essentially acts as a ``stretch" parameter in the time dimension, with slower values of $v_{\rm max}$ producing slower light curve evolution at all stages. 

By applying observational constraints on two of the four fundamental parameters with measurements, we can find non-degenerate fits to the other two; for the remainder of our discussion we consider $v_{\rm max}$ and $d$ to be constrained by optical spectroscopy and light echo imaging with the {\it HST} (Figure \ref{hbeta}, \citealt{Osborne11}, Sokoloski et al. 2013).  We find that the ejected mass is in the range $(1-3) \times 10^{-4}$ M$_{\odot}$ allowing for the 0.5 kpc uncertainty on the light echo-derived distance and assuming the ejecta velocity is in the range 2,000--4,000 km s$^{-1}$.  The best fit models for all parameter combinations have delays of $\sim$75 days in the ejection time relative to the optical discovery, showing that this result is driven primarily by the early non-detections of the source, and is independent of the choice of $v_{\rm max}$ or $d$.  The thickness of the shell, $v_{\rm min}/v_{\rm max}$, is largely determined by how flat-topped the radio light curve is, and is also independent of the assumed values of $v_{\rm max}$ or $d$.  

\subsection{Effects of geometry on the derived ejecta parameters }
In the simple modeling applied thus far, we assumed that the ejecta can be described as a smooth, spherical shell of gas. This picture is undoubtedly overly-simplistic.   A detailed examination of the optical line evolution during the 2011 outburst suggests that even at early times, the ejecta in T~Pyx are bipolar in morphology \citep{Chesneau11, Shore13}.  Studies of the radio evolution of bipolar ejecta have shown that accounting for non-spherical geometry can change the mass estimates derived from radio light curves, with the mass value dropping as the ejecta become more conical  \citep{Seaquist77}.  Accounting for bipolar geometry could reduce our estimated mass by as much as a factor of two (and increase the electron temperature by as much as a factor of four; V.\ Ribeiro et al.\ 2013, in preparation). 

In addition, there is a wealth of evidence from both optical spectroscopy during nova outbursts and imaging of nova remnants that ejecta are often clumpy, with volume filling factors in the range $f = 10^{-5}-10^{-1}$ \citep[e.g.,][]{Williams94, Andrea_etal94, Saizar_Ferland94, Mason_etal05, Ederoclite_etal06, Shara_etal12}. T~Pyx itself shows significant evidence for clumping in its older ejecta, as the surrounding H$\alpha$+[\ion{N}{2}] nebula has been resolved into a ring of knots by \emph{HST} \citep{Shara_etal97, Schaefer10a}. 

Allowing for clumping in the ejecta will impact our derived ejected mass.   We can adapt our simple model of radio emission from novae to account for clumping if we assume that a single volume filling factor ($f$) can describe the ejecta at all times and radii, and that that the material between clumps is of zero density. We also continue to assume that the ejecta are isothermal. In this case, clumping simply increases the optical depth by a constant factor, $1/f$, over the uniform-density case \citep{Abbott_etal81}. Clumpiness therefore has the effect of prolonging the optically-thick phase and increasing the flux density during the partially optically thick and optically-thin phases.
Accounting for this flux boosting, the best-fits to the data with clumpy shell models imply an ejected mass in T~Pyx of
\begin{equation}
M_{ej} = 2 \times 10^{-4}~ f^{0.5} ~{\rm M_{\odot}}.
\end{equation}
The electron temperature derived from model fits is largely independent of the assumed filling factor.
\citet{Shore11} estimate a filling factor, $f = 0.03$, by modeling the optical spectrum of T~Pyx on Day 170.  With this value of $f$, our best-fit clumpy ejecta model yields a mass of $4 \times 10^{-5}$ M$_{\odot}$.  Taking into account the additional factor of two reduction for a bipolar ejection, and the factor of two uncertainty from the distance and expansion velocity, the ejected mass could be as low as $1 \times 10^{-5}$ M$_{\odot}$. 

We note, however, that non-spherical and/or non-uniform geometries cannot account for either the delayed, complicated radio rise we have observed in T~Pyx, or the steep decline phase.  Allowing for clumpy ejecta simply results in a scaling of a radio light curve, but does not fundamentally alter its shape. Neither does bipolar geometry offer an explanation for the observed flux evolution.  
Regardless of the geometry, we still expect a flux density that increases as $t^{2}$ while the ejecta are optically thick, as long as they are in free expansion.  As we discussed in Section 3.2, such a light curve rise is not observed in T~Pyx.  Furthermore, bipolar models decline at the expected $t^{-3}$ rate, and cannot account for the steep decline we see in the radio light curve after Day 200.

\section{Alternative scenarios for producing radio emission in T~Pyx}

We assert in the previous section that the radio light curve is the product of material ejected during the 2011 nova outburst. In this section we show that other origins for the radio light curve can be confidently ruled out. 

\subsection{Radio emission from the central binary?}
The radio flux densities observed with the VLA on the optically-thick rise rule out the central binary as the source of radio emission.  In quiescence, the radio emission from T~Pyx is extremely faint.  Observations obtained with the VLA in 2000 place 3-$\sigma$ upper limits on the flux densities at 4.8 and 8.4 GHz of 66 and 57 $\mu$Jy, respectively \citep{Ogley02}.  The flux densities observed during the outburst are much brighter than this.  Since the steep spectrum of the rising part of the radio light curve is consistent with optically-thick thermal emission, we can make an estimate of the size of the emitting region if we assume an electron temperature and a distance to the source (Equation 1).  
The flux densities observed at 5 and 7 GHz on Day 66 imply an angular diameter of the source of 0.008 arcsec assuming a conservatively high temperature, $10^{5}$ K (if the electron temperature is lower, the size is larger). At a distance of 4.8 kpc, this angular size translates to a physical radius of the source of $\sim$4 x 10$^{14}$ cm---much larger than the size of the binary system, which is a few 10$^{10}$ cm using the parameters of \citet{Uthas10}.  

An alternative scenario is that the radio emission originates in a dense, pre-existing medium in the immediate vicinity of the central binary. This material would be much closer in than the clumps in the extended optical nebula resolved with {\it HST}.  \citet{Chomiuk_etal12} showed that the radio emission in the recent nova V407~Cyg originated primarily in the dense wind of the Mira giant companion in that system.  During the period where stable nuclear burning took place on the white dwarf's surface, the circumbinary medium was illuminated and ionized, and the radio flux density from this material increased.  Could a similar mechanism be at play in T~Pyx?  Long-term optical light curves of T~Pyx show a change in the orbital period with time, consistent with mass loss rate from the companion star as high as 5 $\times$ 10$^{-7}$  M$_{\odot}$ yr$^{-1}$ \citep{Patterson98, Uthas10, Schaefer_etal11}. If mass transfer in the system is even mildly non-conservative, then a significant amount of material could be lost into the circumbinary environment, which might emit in the radio once it is ionized by the nova outburst.  

However, radio emission from this circumbinary material is expected to emit in a manner similar to a stellar wind (see discussion in \citealt{Chomiuk_etal12}) and therefore exhibit a partially optically thin spectrum with $\alpha \approx 0.6$.  The spectrum we observe during the rise is more consistent with a completely optically-thick body ($\alpha \approx 1.8$), and the increase in flux density between Days 67 and 174 implies that this body is expanding. Furthermore, if the circumbinary medium is in the immediate vicinity of the white dwarf, it would be difficult to account for the delay of $\gtrsim$50 days between the optical rise and the radio rise---the radio emission should either have appeared immediately if flash ionized by the thermonuclear runaway, or after Day 120 when the soft X-ray emission appears in the {\it Swift} light curve \citep{Chomiuk13}. Therefore, it is unlikely that T~Pyx's radio light curve is the result of  ionization or interaction in a pre-existing circumbinary medium.   

\subsection{Radio emission from the extended remnant?}
Given the extended nebular remnant around T~Pyx visible in H$\alpha$+[\ion{N}{2}], it is conceivable that the radio emission could originate from this spatially-extended material, as the radiation associated with the outburst interacts with (and presumably ionizes) the remnant.  Indeed, the timescales on which we see the radio emission develop are appropriate.  Based on the {\it HST} imaging presented in {\citet{Schaefer10a}, the majority of clumps in the remnant are located at distances between 2.3 and 5.5 arcsec from the central source.  Placing T~Pyx at 4.8 kpc, these separations correspond to physical distances of $(1.6-3.9) \times 10^{17}$ cm from the central system.  The light crossing times to reach the inner and outer extents of the shell are therefore 62 and 149 days, respectively---corresponding roughly to the time we see the radio light curve rise and then fade.

However, the H${\alpha}$ luminosities observed during the outburst from the extended remnant are much too low for the photoionized gas in the nebula to have produced the high observed radio flux density.  An {\it HST} observation of T~Pyx on 2011 September 25 (Day 164) using the F656N filter yields an integrated H$\alpha$ flux for all knots beyond 3.5 arcsec of $\sim$4 $\times$ 10$^{-17}$ erg s$^{-1}$ cm$^{-2}$ (A. Crotts, private communication).  Correcting for an interstellar E(B-V) of 0.49 using the extinction law of \citet{Fitzpatrick99}, and using the conversion relation between H$\alpha$ and 5 GHz fluxes expected for a 10$^{4}$ K plasma given in \citet{Caplan86}, the observed H$\alpha$ flux implies a radio flux density at 5 GHz from the extended remnant of $\sim$0.1 $\mu$Jy---five orders of magnitude lower than observed.  

Moreover, we never resolve the radio source associated with T~Pyx, despite the fact that the VLA was in its extended high-resolution A- and B-configurations around light curve maximum. A nebula with a diameter of a few arcseconds would have been thoroughly sampled with $\gtrsim$10 synthesized beams across it at our higher observing frequencies, but T~Pyx was consistent with a point-like source in all observations. 

Given these factors, and the relatively good match between the multi-frequency light curves and the simple model of expanding thermal ejecta presented in Section \ref{model}, we conclude that the expanding nova ejecta are the most likely origin of the radio emission.  

\section{Discussion}

The ejected masses derived from our model fits to the radio light curve are large compared to expectations for recurrent novae, ranging from 1 $\times$ 10$^{-5}$ M$_{\odot}$ for clumped, bipolar ejecta, to 3 $\times$ 10$^{-4}$ M$_{\odot}$ for a shell with a smooth spherical density distribution.  Despite some uncertainties in the geometry, density profile, and temperature evolution of the ejecta, the primary constraint on the ejecta mass is simply provided by the high flux density observed at light curve maximum.  \citet{Selvelli08} estimated an ejecta mass in the range 10$^{-5}$--10$^{-4}$ M$_{\odot}$ for the 1966 outburst based on the long duration of the optically thick phase in the optical light curve---the radio data presented here support that conclusion.  Furthermore, \citet{Patterson13} find a significant slow-down in the orbital period modulation observed in optical photometry of T~Pyx after the 2011 outburst.  They interpret this as the ejection of at least $3 \times 10^{-5}$ M$_{\odot}$ of material from the system during the outburst, in excellent agreement with the range of ejecta masses we derive from the radio data.  Note that the Patterson et al.~ result is independent of the distance to the system.  This high ejected mass challenges commonly held assumptions about T~Pyx and holds important implications for the global parameters of the system. 

\subsection{A high quiescent accretion rate and secular evolution of the white dwarf}
Since we know the recurrence time of T~Pyx, we can estimate the accretion rate in the system.  Assuming that all the material ejected in the 2011 outburst has been accreted since the 1966 outburst, the average accretion rate is simply given by $\dot{M} = \frac{M_{\rm ej}}{t_{\rm rec}}$.  For a recurrence time $t_{\rm rec} = 44$ years, our estimated range of $M_{\rm ej}$ implies an accretion rate in the range $\dot{M} = (0.2-7) \times 10^{-6}\ M_{\odot}$~yr$^{-1}$.  

There are a few estimates of the accretion rate in T~Pyx based on observations.  \citet{Patterson98} and \citet{Uthas10} present determinations of the orbital period of T~Pyx and find significant evidence of a period slow-down, implying mass loss from the companion star. \citet{Uthas10} derive a mass loss rate of $5 \times 10^{-7}$ M$_{\odot}$~yr$^{-1}$ using their estimate of the mass ratio in the binary, and note that this is an upper limit on the mass accretion rate onto the white dwarf (the rate could be lower if mass transfer is non-conservative).  \citet{Selvelli08} use ultraviolet observations of T~Pyx in quiescence to estimate an accretion rate of $(1.1 \pm 0.3) \times 10^{-8}$ M$_{\odot}$ yr$^{-1}$, seemingly much lower than the values we infer from the radio data.  However, if we allow for greater reddening, a lower mass white dwarf, and the updated distance to T Pyx of 4.8 kpc, then the UV flux reported in \citet{Selvelli08} is consistent with a mass accretion rate as high as $\sim$10$^{-7}$ M$_{\odot}$~yr$^{-1}$ (a factor of 10 larger than originally reported). This would bring the accreted mass in line with the low end of our ejected mass estimate. Most recently, \citet{Godon14} estimate the accretion rate based on accretion disk model fits to the archival {\it IUE} data in combination with {\it GALEX} and {\it HST} spectra.  They estimate an accretion rate as high as 2.7 $\times$ 10$^{-6}$ M$_{\odot}$ yr$^{-1}$; significantly higher than previous estimates.  This implies an accreted mass as high as 10$^{-4}$ M$_{\odot}$, although this result depends on both reddening and white dwarf mass assumed, as well as the inclination of the binary system.  

If the lower accretion rate estimates are correct, then we must allow for some dredge up of the white dwarf material to bring the measurements of ejected mass and accretion rate into better agreement.  Theoretical studies of nova outbursts find that some novae eject as much as three times the accreted mass \citep{Yaron_etal05}, in which case the  accretion rate could be up to a factor of $\sim$3 lower than the value estimated above from our ejecta mass measurement (i.e. $8 \times 10^{-8} - 2 \times 10^{-6}$ M$_{\odot}$ yr$^{-1}$).  Accretion rates on the low end of this range can produce novae with recurrence time of 40--50 years if the white dwarf is well below the Chandrasekhar mass ($\sim$1.0--1.2 M$_{\odot}$; \citealt{Wolf13}). Note that the ejected shells for novae on lower mass white dwarfs are expected to be more massive, in line with the estimates we derive here from the radio light curve.  A net mass-loss with each outburst, combined with a low mass white dwarf, would suggest that T Pyx is not a Type Ia supernova progenitor.

If, on the other hand, the more recent estimate of a high accretion rate in quiescence is correct and representative of the average state of the system, then T Pyx could have comfortably accreted the mass necessary  to account for observed radio flux since the previous outburst.  Again, this scenario favors a low mass white dwarf, since the recurrence time should have been much shorter for a massive white dwarf accreting at these rates  \citep{Yaron_etal05}.  At these high accretion rates, there are predictions that the envelope should either puff up to red giant proportions \citep[e.g.][]{Iben84} or power steady nuclear burning \citep[e.g.][]{Nomoto07}.  The quiescence observations of T Pyx are inconsistent with either of these pictures \citet{Selvelli08}.  More recent studies have found that nova outbursts can happen on white dwarfs accreting at much higher rates than previous studies \citep{Starrfield12,Idan13}.  T Pyx may be an example of such a system, although the high ejected mass is not found for most outbursts in these models.

\subsection{Complex mass ejection physics in T~Pyx}
As we showed in Section 3.2, the early part of the radio light curve of T~Pyx is inconsistent with a shell expanding from the time of discovery at constant velocity or temperature. We require dramatic changes in either the temperature or velocity (or both) of the ejected shell over the first $\sim$100 days of the outburst.  Temperature changes in the ejecta are possibly more straightforward to understand---the ejected shell can cool quickly via adiabatic expansion at early times, and then get reheated later once it becomes optically-thin to the hot radiation from the central nuclear-burning white dwarf.  The temperatures we have derived from the radio flux density limits are reasonable if the expansion is fast, and could be achieved early in the outburst.  Even if material closer to the white dwarf is warm, as long as the outer ejecta are shielded from the radiation field of the white dwarf and remain cold, the radio photosphere would reflect this outer temperature. 

Changes in expansion velocity are more difficult to explain in the context of a nova outburst.  The low velocities at early times could be due to stalled expansion of the nova shell on an $\sim$AU scale, followed by a later epoch of rapid expansion, perhaps through common envelope evolution \citep[see e.g.,][]{Livio90}.  Such scenarios have been proposed to explain the long optical plateaus seen in some nova \citep{Friedjung92, Kato11}, and could apply to T~Pyx given the prolonged plateau observed in the outburst light curve.   However, this scenario alone requires a very sudden and dramatic increase in the ejection velocity between days 48 and 66 (see dashed lines in lower panel of Figure \ref{thickrad}). Some combination of a delayed ejection and temperature evolution of the ejecta would imply a more gradual change in expansion velocity. 

Multi-wavelength observations hold the key to determining which of the two scenarios is correct.  Emission line information, such as that shown in Figure \ref{hbeta}, certainly show evidence for faster material at late times.   However, very cold material would probably not be apparent in the optical data, since such cold electrons would not excite the relevant spectral transitions.  X-ray observations can play an important role, since they trace (1) the appearance of  the hot white dwarf surface once the ejecta have expanded sufficiently, and (2) any shock interactions between discrete shells of material. Indeed, there is evidence for both types of X-ray emission in T~Pyx \citep{Tofflemire13,Surina14}, and the relative timing of their appearance holds information about the presence of a delayed ejection of material from the vicinity of the white dwarf.  We explore the X-ray light curve and its implication for mass ejection in T~Pyx in a companion paper, \citet{Chomiuk14}.  

\section{Conclusions}
From our analysis of multi-frequency radio light curves obtained with the newly-upgraded VLA, we reach several surprising conclusions about mass ejection during the 2011 outburst of T~Pyx.
\begin{enumerate}
\item The 2011 nova event in T~Pyx expelled a large amount of mass, totaling $(1-30) \times 10^{-5}\ M_{\odot}$ for a very conservative set of assumptions about distance to the nova, maximum ejecta velocity, and the geometry of the ejecta. 

\item If all the radio emitting material was ejected from the white dwarf surface during this nova outburst, then only the highest estimates of the accretion rate in quiescence allow enough material to be accreted since the previous outburst.  Otherwise, the large ejected mass implies that some white dwarf material was dredged up during the outburst.  


\item Both the ejected mass and the inferred accretion rate in quiescence suggest that the white dwarf in T~Pyx is not near the Chandrasekhar mass. Instead, it more likely has a mass of 1--1.2 $M_{\odot}$.  

\item The nova shell was not ejected in a simple, impulsive event.  The ejecta experienced either a dramatic cooling and reheating, a delayed expulsion, or some combination of the two. A full investigation of the multi-wavelength evolution of this outburst of T~Pyx will be required to fully characterize the manner in which mass was ejected from the system. 
\end{enumerate}

\acknowledgements
We are grateful to H.~Uthas, R.~Williams, M.~Shara, J.~Patterson, S.~Starrfield, M.\ Kato and D.\ Prialnik for illuminating discussions, and to the anonymous referee for their feedback on this work. We thank NRAO for the generous allocation of director's discretionary time that made this work possible. We are also grateful to the VLA commissioning team, including J.~McMullin, J.~Wrobel, E.~Momjian, L.~Sjouwerman, and G.~van Moorsel, for their assistance in the acquisition of this data set. The National Radio Astronomy Observatory is a facility of the National Science Foundation operated under cooperative agreement by Associated Universities, Inc. This work was carried out while L.~Chomiuk and N.~Roy were Jansky Fellows of the National Radio Astronomy Observatory. J.~L.~S. and J.~W. acknowledge support from the National Science Foundation through award AST-1217778.  Finally, we acknowledge with thanks the variable star observations from the AAVSO International Database contributed by observers worldwide and used in this research.\\

{\it Facilities:} \facility{AAVSO}, \facility{Karl G. Jansky VLA}

\bibliography{tpyx.bib}

\begin{thebibliography}{63}
\expandafter\ifx\csname natexlab\endcsname\relax\def\natexlab#1{#1}\fi

\bibitem[{{Abbott} {et~al.}(1981){Abbott}, {Bieging}, \&
  {Churchwell}}]{Abbott_etal81}
{Abbott}, D.~C., {Bieging}, J.~H., \& {Churchwell}, E. 1981, \apj, 250, 645

\bibitem[{{Andre\"{a}} {et~al.}(1994){Andre\"{a}}, {Drechsel}, \&
  {Starrfield}}]{Andrea_etal94}
{Andre\"{a}}, J., {Drechsel}, H., \& {Starrfield}, S. 1994, \aap, 291, 869

\bibitem[{{Caplan} \& {Deharveng}(1986)}]{Caplan86}
{Caplan}, J., \& {Deharveng}, L. 1986, \aap, 155, 297

\bibitem[{{Chesneau} {et~al.}(2011)}]{Chesneau11}
{Chesneau}, O., {et~al.} 2011, \aap, 534, L11

\bibitem[{{Chomiuk} {et~al.}(2012){Chomiuk}, {Krauss}, {Rupen}, {Nelson},
  {Roy}, {Sokoloski}, {Mukai}, {Munari}, {Mioduszewski}, {Weston}, {O'Brien},
  {Eyres}, \& {Bode}}]{Chomiuk_etal12}
{Chomiuk}, L., {Krauss}, M.~I., {Rupen}, M.~P., {et~al.} 2012, \apj, 761, 173

\bibitem[{{Chomiuk} {et~al.}(2014){Chomiuk}, {Nelson}, {Mukai}, {Sokoloski},
  {Mukai}, {Krauss}, {Mioduszewski}, {Rupen}, \& {Weston}}]{Chomiuk14}
{Chomiuk}, L., {Nelson}, T., {Mukai}, K., {et~al.} 2014, arXiv

\bibitem[{{Ederoclite} {et~al.}(2006){Ederoclite}, {Mason}, {Della Valle},
  {Gilmozzi}, {Williams}, {Germany}, {Saviane}, {Matteucci}, {Schaefer},
  {Walter}, {Rudy}, {Lynch}, {Mazuk}, {Venturini}, {Puetter}, {Perry},
  {Liller}, \& {Rotter}}]{Ederoclite_etal06}
{Ederoclite}, A., {Mason}, E., {Della Valle}, M., {et~al.} 2006, \aap, 459, 875

\bibitem[{{Eyres} {et~al.}(1996){Eyres}, {Davis}, \& {Bode}}]{Eyres96}
{Eyres}, S.~P.~S., {Davis}, R.~J., \& {Bode}, M.~F. 1996, \mnras, 279, 249

\bibitem[{{Fitzpatrick}(1999)}]{Fitzpatrick99}
{Fitzpatrick}, E.~L. 1999, \pasp, 111, 63

\bibitem[{{Friedjung}(1992)}]{Friedjung92}
{Friedjung}, M. 1992, \aap, 262, 487

\bibitem[{{Gehrz} {et~al.}(1998){Gehrz}, {Truran}, {Williams}, \&
  {Starrfield}}]{Gehrz98}
{Gehrz}, R.~D., {Truran}, J.~W., {Williams}, R.~E., \& {Starrfield}, S. 1998,
  \pasp, 110, 3

\bibitem[{{Godon} {et~al.}(2014){Godon}, {Sion}, {Starrfield}, {Livio},
  {Williams}, {Woodward}, {Kuin}, \& {Page}}]{Godon14}
{Godon}, P., {Sion}, E.~M., {Starrfield}, S., {et~al.} 2014, ArXiv e-prints

\bibitem[{{Heywood} {et~al.}(2005){Heywood}, {O'Brien}, {Eyres}, {Bode}, \&
  {Davis}}]{Heywood05}
{Heywood}, I., {O'Brien}, T.~J., {Eyres}, S.~P.~S., {Bode}, M.~F., \& {Davis},
  R.~J. 2005, \mnras, 362, 469

\bibitem[{{Hjellming} {et~al.}(1979){Hjellming}, {Wade}, {Vandenberg}, \&
  {Newell}}]{Hjellming79}
{Hjellming}, R.~M., {Wade}, C.~M., {Vandenberg}, N.~R., \& {Newell}, R.~T.
  1979, \aj, 84, 1619

\bibitem[{{Iben} \& {Tutukov}(1984)}]{Iben84}
{Iben}, Jr., I., \& {Tutukov}, A.~V. 1984, \apjs, 54, 335

\bibitem[{{Idan} {et~al.}(2013){Idan}, {Shaviv}, \& {Shaviv}}]{Idan13}
{Idan}, I., {Shaviv}, N.~J., \& {Shaviv}, G. 2013, \mnras, 433, 2884

\bibitem[{{Kantharia} {et~al.}(2007){Kantharia}, {Anupama}, {Prabhu}, {Ramya},
  {Bode}, {Eyres}, \& {O'Brien}}]{Kantharia07}
{Kantharia}, N.~G., {Anupama}, G.~C., {Prabhu}, T.~P., {et~al.} 2007, \apjl,
  667, L171

\bibitem[{{Kato} \& {Hachisu}(2011)}]{Kato11}
{Kato}, M., \& {Hachisu}, I. 2011, \apj, 743, 157

\bibitem[{{Knigge} {et~al.}(2011){Knigge}, {Baraffe}, \&
  {Patterson}}]{Knigge11}
{Knigge}, C., {Baraffe}, I., \& {Patterson}, J. 2011, \apjs, 194, 28

\bibitem[{{Knigge} {et~al.}(2000){Knigge}, {King}, \& {Patterson}}]{Knigge00}
{Knigge}, C., {King}, A.~R., \& {Patterson}, J. 2000, \aap, 364, L75

\bibitem[{{Kwok}(1983)}]{Kwok83}
{Kwok}, S. 1983, \mnras, 202, 1149

\bibitem[{{Lang}(1980)}]{Lang80}
{Lang}, K.~R. 1980, {Astrophysical Formulae. A Compendium for the Physicist and
  Astrophysicist.} (Springer-Verlag: Berlin)

\bibitem[{{Livio}(2001)}]{Livio01}
{Livio}, M. 2001, in Supernovae and Gamma-Ray Bursts: the Greatest Explosions
  since the Big Bang, ed. M.~{Livio}, N.~{Panagia}, \& K.~{Sahu}, 334--355

\bibitem[{{Livio} {et~al.}(1990){Livio}, {Shankar}, {Burkert}, \&
  {Truran}}]{Livio90}
{Livio}, M., {Shankar}, A., {Burkert}, A., \& {Truran}, J.~W. 1990, \apj, 356,
  250

\bibitem[{{MacDonald}(1984)}]{MacDonald84}
{MacDonald}, J. 1984, \apj, 283, 241

\bibitem[{{Maoz} {et~al.}(2014){Maoz}, {Mannucci}, \& {Nelemans}}]{Maoz14}
{Maoz}, D., {Mannucci}, F., \& {Nelemans}, G. 2014, \araa

\bibitem[{{Markwardt}(2009)}]{Markwardt09}
{Markwardt}, C.~B. 2009, in Astronomical Data Analysis Software and Systems
  XVIII, ed. D.~A. {Bohlender}, D.~{Durand}, \& P.~{Dowler}, Vol. 411, 251

\bibitem[{{Mason} {et~al.}(2005){Mason}, {Della Valle}, {Gilmozzi}, {Lo Curto},
  \& {Williams}}]{Mason_etal05}
{Mason}, E., {Della Valle}, M., {Gilmozzi}, R., {Lo Curto}, G., \& {Williams},
  R.~E. 2005, \aap, 435, 1031

\bibitem[{{Nomoto} {et~al.}(2007){Nomoto}, {Saio}, {Kato}, \&
  {Hachisu}}]{Nomoto07}
{Nomoto}, K., {Saio}, H., {Kato}, M., \& {Hachisu}, I. 2007, \apj, 663, 1269

\bibitem[{{Ogley} {et~al.}(2002){Ogley}, {Chaty}, {Crocker},
  {et~al.}}]{Ogley02}
{Ogley}, R.~N., {Chaty}, S., {Crocker}, M., {et~al.} 2002, \mnras, 330, 772

\bibitem[{{Osborne} {et~al.}(2011){Osborne}, {Beardmore}, {Page},
  {et~al.}}]{Osborne11}
{Osborne}, J.~P., {Beardmore}, A.~P., {Page}, K.~L., {et~al.} 2011, ATel, 3549

\bibitem[{{Panagia} \& {Felli}(1975)}]{Panagia_Felli75}
{Panagia}, N., \& {Felli}, M. 1975, \aap, 39, 1

\bibitem[{{Patterson} {et~al.}(1998){Patterson}, {Kemp}, {Shambrook},
  {Thorstensen}, {et~al.}}]{Patterson98}
{Patterson}, J., {Kemp}, J., {Shambrook}, A., {Thorstensen}, {et~al.} 1998,
  \pasp, 110, 380

\bibitem[{{Patterson} {et~al.}(2013){Patterson}, {Oksanen}, {Monard},
  {et~al.}}]{Patterson13}
{Patterson}, J., {Oksanen}, A., {Monard}, B., {et~al.} 2013, arXiv 1303.0736

\bibitem[{{Pavelin} {et~al.}(1993){Pavelin}, {Davis}, {Morrison}, {Bode}, \&
  {Ivison}}]{Pavelin93}
{Pavelin}, P.~E., {Davis}, R.~J., {Morrison}, L.~V., {Bode}, M.~F., \&
  {Ivison}, R.~J. 1993, \nat, 363, 424

\bibitem[{{Rupen} {et~al.}(2008){Rupen}, {Mioduszewski}, \&
  {Sokoloski}}]{Rupen08}
{Rupen}, M.~P., {Mioduszewski}, A.~J., \& {Sokoloski}, J.~L. 2008, \apj, 688,
  559

\bibitem[{{Saizar} \& {Ferland}(1994)}]{Saizar_Ferland94}
{Saizar}, P., \& {Ferland}, G.~J. 1994, \apj, 425, 755

\bibitem[{{Schaefer}(2010)}]{Schaefer10b}
{Schaefer}, B.~E. 2010, \apjs, 187, 275

\bibitem[{{Schaefer} {et~al.}(1992){Schaefer}, {Landolt}, {Vogt}, {Buckley},
  {Warner}, {Walker}, \& {Bond}}]{Schaefer92}
{Schaefer}, B.~E., {Landolt}, A.~U., {Vogt}, N., {et~al.} 1992, \apjs, 81, 321

\bibitem[{{Schaefer} {et~al.}(2010){Schaefer}, {Pagnotta}, \&
  {Shara}}]{Schaefer10a}
{Schaefer}, B.~E., {Pagnotta}, A., \& {Shara}, M.~M. 2010, \apj, 708, 381

\bibitem[{{Schaefer} {et~al.}(2013){Schaefer}, {Landolt}, {Linnolt},
  {Stubbings}, {Pojmanski}, {Plummer}, {Kerr}, {Nelson}, {Carstens},
  {Streamer}, {Richards}, {Myers}, \& {Dillon}}]{Schaefer_etal11}
{Schaefer}, B.~E., {Landolt}, A.~U., {Linnolt}, M., {et~al.} 2013, \apj, 773,
  55

\bibitem[{{Schatzman}(1949)}]{Schatzman49}
{Schatzman}, E. 1949, Annales d'Astrophysique, 12, 281

\bibitem[{{Seaquist} \& {Bode}(2008)}]{Seaquist_Bode08}
{Seaquist}, E.~R., \& {Bode}, M.~F. 2008, in {Classical Novae, 2nd
  Edition.~Cambridge Astrophysics Series, No.~43, Cambridge: Cambridge
  University Press}, ed. {M.~F.~Bode \& A.~Evans}, 141

\bibitem[{{Seaquist} {et~al.}(1980){Seaquist}, {Duric}, {Israel}, {Spoelstra},
  {Ulich}, \& {Gregory}}]{Seaquist80}
{Seaquist}, E.~R., {Duric}, N., {Israel}, F.~P., {et~al.} 1980, \aj, 85, 283

\bibitem[{{Seaquist} \& {Palimaka}(1977)}]{Seaquist77}
{Seaquist}, E.~R., \& {Palimaka}, J. 1977, \apj, 217, 781

\bibitem[{{Selvelli} {et~al.}(2008){Selvelli}, {Cassatella}, {Gilmozzi}, \&
  {Gonz{\'a}lez-Riestra}}]{Selvelli08}
{Selvelli}, P., {Cassatella}, A., {Gilmozzi}, R., \& {Gonz{\'a}lez-Riestra}, R.
  2008, \aap, 492, 787

\bibitem[{{Shara} {et~al.}(2012){Shara}, {Zurek}, {De Marco}, {Mizusawa},
  {Williams}, \& {Livio}}]{Shara_etal12}
{Shara}, M.~M., {Zurek}, D., {De Marco}, O., {et~al.} 2012, \aj, 143, 143

\bibitem[{{Shara} {et~al.}(1997){Shara}, {Zurek}, {Williams}, {Prialnik},
  {Gilmozzi}, \& {Moffat}}]{Shara_etal97}
{Shara}, M.~M., {Zurek}, D.~R., {Williams}, R.~E., {et~al.} 1997, \aj, 114, 258

\bibitem[{{Shore} {et~al.}(2011){Shore}, {Augusteijn}, {Ederoclite}, \&
  {Uthas}}]{Shore11}
{Shore}, S.~N., {Augusteijn}, T., {Ederoclite}, A., \& {Uthas}, H. 2011, \aap,
  533, L8

\bibitem[{{Shore} {et~al.}(2013){Shore}, {Schwarz}, {De Gennaro Aquino},
  {Augusteijn}, {Walter}, {Starrfield}, \& {Sion}}]{Shore13}
{Shore}, S.~N., {Schwarz}, G.~J., {De Gennaro Aquino}, I., {et~al.} 2013, \aap,
  549, A140

\bibitem[{{Sokoloski} {et~al.}(2013){Sokoloski}, {Crotts}, {Lawrence}, \&
  {Uthas}}]{Sokoloski13}
{Sokoloski}, J.~L., {Crotts}, A.~P.~S., {Lawrence}, S., \& {Uthas}, H. 2013,
  \apjl, 770, L33

\bibitem[{{Starrfield} {et~al.}(2012){Starrfield}, {Timmes}, {Iliadis}, {Hix},
  {Arnett}, {Meakin}, \& {Sparks}}]{Starrfield12}
{Starrfield}, S., {Timmes}, F.~X., {Iliadis}, C., {et~al.} 2012, Baltic
  Astronomy, 21, 76

\bibitem[{{Starrfield} {et~al.}(1972){Starrfield}, {Truran}, {Sparks}, \&
  {Kutter}}]{Starrfield72}
{Starrfield}, S., {Truran}, J.~W., {Sparks}, W.~M., \& {Kutter}, G.~S. 1972,
  \apj, 176, 169

\bibitem[{{Surina} {et~al.}(2014){Surina}, {Hounsell}, {Bode}, {Darnley},
  {Harman}, \& {Walter}}]{Surina14}
{Surina}, F., {Hounsell}, R.~A., {Bode}, M.~F., {et~al.} 2014, ArXiv e-prints

\bibitem[{{Taylor} {et~al.}(1988){Taylor}, {Hjellming}, {Seaquist}, \&
  {Gehrz}}]{Taylor88}
{Taylor}, A.~R., {Hjellming}, R.~M., {Seaquist}, E.~R., \& {Gehrz}, R.~D. 1988,
  \nat, 335, 235

\bibitem[{{Tofflemire} {et~al.}(2013){Tofflemire}, {Orio}, {Page}, {Osborne},
  {Ciroi}, {Cracco}, {Di Mille}, \& {Maxwell}}]{Tofflemire13}
{Tofflemire}, B.~M., {Orio}, M., {Page}, K.~L., {et~al.} 2013, \apj, 779, 22

\bibitem[{{Uthas} {et~al.}(2010){Uthas}, {Knigge}, \& {Steeghs}}]{Uthas10}
{Uthas}, H., {Knigge}, C., \& {Steeghs}, D. 2010, \mnras, 409, 237

\bibitem[{{Waagan} {et~al.}(2011){Waagan}, {Linnolt}, {Bolzoni}, {Amorim},
  {Plummer}, {Williams}, {Matsuyama}, {Kerr}, \& {Pearce}}]{Waagan_etal11}
{Waagan}, E., {Linnolt}, M., {Bolzoni}, S., {et~al.} 2011, CBET, 2700

\bibitem[{{Walter} {et~al.}(2012){Walter}, {Battisti}, {Towers}, {Bond}, \&
  {Stringfellow}}]{Walter12}
{Walter}, F.~M., {Battisti}, A., {Towers}, S.~E., {Bond}, H.~E., \&
  {Stringfellow}, G.~S. 2012, \pasp, 124, 1057

\bibitem[{{Williams}(1982)}]{Williams82}
{Williams}, R.~E. 1982, \apj, 261, 170

\bibitem[{{Williams}(1994)}]{Williams94}
---. 1994, \apj, 426, 279

\bibitem[{{Wolf} {et~al.}(2013){Wolf}, {Bildsten}, {Brooks}, \&
  {Paxton}}]{Wolf13}
{Wolf}, W.~M., {Bildsten}, L., {Brooks}, J., \& {Paxton}, B. 2013, arXiv
  1309.3375

\bibitem[{{Yaron} {et~al.}(2005){Yaron}, {Prialnik}, {Shara}, \&
  {Kovetz}}]{Yaron_etal05}
{Yaron}, O., {Prialnik}, D., {Shara}, M.~M., \& {Kovetz}, A. 2005, \apj, 623,
  398

\end{thebibliography}

\newpage
\begin{deluxetable}{cccccccccccccc}
\tablewidth{0 pt}
\tabletypesize{\footnotesize}
\setlength{\tabcolsep}{0.025in}
\tablecaption{ \label{tab:phot}
VLA Observations of T~Pyx }
\tablehead{Date & $t-t_{0}\tablenotemark{a}$ & Epoch  & Conf. & Freq & $S_\nu$ & Freq & $S_\nu$ & Freq & $S_\nu$ & Freq & $S_\nu$ \\
(UT) & (Days) & & & (GHz) & (mJy) & (GHz) & (mJy) & (GHz) & (mJy) & (GHz) & (mJy)}
\startdata
2011 Apr 22.0 & 8.0 & 1 & B & 5.9 & $0.003\pm0.008$ & & & & & & \\
2011 Apr 22.2 & 8.2 & 1 & B & 33.1 & $0.016\pm0.030$ & & &  & & & \\
 &  &  &  & & & & & & & & \\

2011 Apr 30.9 & 16.9 & 2 & B & 5.9 & $0.027\pm0.011$&  & & & & & \\
2011 May 1.1 & 17.1 & 2 & B & 33.1 & $0.248\pm0.040$ &  & &  & & & \\
 &  &  &  & & & & & & & & \\

2011 May 15.5 & 31.5 & 3 & BnA & 1.6 & $-0.040\pm0.031$ & 6.0 & $0.022\pm0.006$ & 14.3 & $0.028\pm0.017$& & \\
2011 May 15.9 & 31.9 & 3 & BnA & 3.0 & $0.000\pm0.035$ & 8.4 & $0.017\pm0.018$& 22.0 & $0.038\pm0.049$ & 33.0 & $-0.046\pm0.055$ \\
 &  &  &  & & & & & & & & \\

2011 May 29.0 & 45.0 & 4 & BnA & 1.5 & $0.134\pm0.068$ & 3.0 & $0.015\pm0.034$ & 6.0 & $0.005\pm0.010$ & 8.4 & $0.032\pm0.017$ \\
 & &  & & 14.3 & $0.026\pm0.019$ & 22.0 & $0.021\pm0.031$& 33.0 & $0.011\pm0.030$ & & \\
 &  &  &  & & & & & & & & \\

2011 Jun 1.0 & 48.0 & 5 & BnA$\rightarrow$A & 33.1 & $0.048\pm0.029$ & & &  & & & \\
 &  &  &  & & & & & & & & \\

2011 Jun 19.9 & 66.9 & 6 & A & 5.0 & $0.130\pm0.017$ & 7.0 & $0.194\pm0.016$& & & & \\
 &  &  &  & & & & & & & & \\

2011 Jul 24.8 & 101.8 & 7 & A & 1.6 & $0.082\pm0.040$ & 2.5 & $0.119\pm0.052$ & 3.5 & $0.185\pm0.039$ & 5.0 & $0.304\pm0.031$\\
 & &  & & 7.0 & $0.529\pm0.039$ & & &  & & & \\
 &  &  &  & & & & & & & & \\

2011 Aug 20.7 & 128.7 & 8 & A & 1.3 & $0.146\pm0.058$ & 1.8 & $0.190\pm0.066$ & 2.5 & $0.318\pm0.056$ & 3.5 & $0.581\pm0.065$\\
 & &  & & 5.0 & $1.128\pm0.061$ & 7.0 & $1.885\pm0.097$ & 8.4 & $2.809\pm0.141$ & 12.6 & $4.832\pm0.500$\\
 & &  & & 16.0 & $6.932\pm0.715$ & 22.0 & $11.38\pm1.17$ & 29.0 & $18.82\pm1.91$ & 37.0 & $23.28\pm2.36$\\
 &  &  &  & & & & & & & & \\

2011 Sep 17.6 & 156.6 & 9 &  A$\rightarrow$D & 12.6 & $11.09\pm0.12$ & 16.0 & $14.20\pm1.44$ & 23.7 & $26.93\pm2.71$ & 29.0 & $38.97\pm3.94$\\
 & &  & & 37.0 & $47.90\pm4.94$ & & &  & & & \\
 &  &  &  & & & & & & & & \\

2011 Sep 24.6 & 163.6 & 10 & A$\rightarrow$D & 2.5 & $0.800\pm0.126$ & 3.5 & $1.350\pm0.098$ & 5.0 & $3.014\pm0.161$ & 7.0 & $5.157\pm0.260$\\
 & &  & & 8.4 & $6.738\pm0.338$ & & &  & & & \\
 &  &  &  & & & & & & & & \\
 
2011 Oct 25.5 & 194.5 & 11 & D & 1.3 & $-0.763\pm0.210$ & 1.8 & $0.631\pm0.210$ & 2.5 & $1.421\pm0.139$ & 3.5 & $1.716\pm0.127$\\
 & &  & & 5.0 &  $3.816\pm0.196$ & 7.0 & $6.106\pm0.308$ & 12.9 & $14.09\pm1.41$ & 16.0 & $18.60\pm1.86$ \\
  & &  & & 23.7 & $26.27\pm2.63$ & 29.0 & $30.97\pm3.10$ & 37.0 & $34.68\pm3.47$ & & \\
 &  &  &  & & & & & & & & \\

2011 Nov 10.5 & 210.5 & 12 & D & 1.3 & $ 0.057\pm0.243$ & 1.8 & $ 0.774\pm0.243$ & 2.5 & $ 1.166\pm0.136$ & 3.5 & $ 1.797\pm0.124$ \\
 & &  & & 5.0 & $ 3.685\pm0.190$ & 7.0 & $ 6.263\pm0.315$ & 12.6 & $14.16\pm1.42$ & 16.0 & $17.64\pm1.77$ \\
 & &  & & 23.7 & $26.32\pm2.63$ & 29.0 & $30.51\pm3.05$ & 37.0 & $33.40\pm3.35$ & & \\
 &  &  &  & & & & & & & & \\

2011 Dec 7.4 & 237.4 & 13 & D & 1.3 & $ 0.699\pm0.277$ & 1.8 & $ 0.445\pm0.167$ &  2.5 & $ 1.567\pm0.121$ & 3.5 & $ 2.307\pm0.128$ \\
 & &  & &  5.0 & $ 4.270\pm0.219$ & 7.0 & $ 6.594\pm0.332$ & 13.3 & $12.41\pm1.25$ &16.0 & $14.37\pm1.44$ \\
 & &  & & 23.7 & $18.14\pm1.82$ & 29.0 & $18.87\pm1.89$ & 37.0 & $19.80\pm1.99$ & & \\
 &  &  &  & & & & & & & & \\

2012 Jan 5.4 & 266.4 & 14 & DnC & 1.3 & $ 0.277\pm0.210$ & 1.8 & $ 0.477\pm0.121$ & 2.5 & $ 1.528\pm0.146$ & 3.5 & $ 2.659\pm0.153$ \\
 & &  & &  5.0 & $ 4.345\pm0.222$ & 7.0 & $ 6.096\pm0.307$ &13.3 & $ 8.995\pm0.903$ & 16.0 & $ 9.434\pm0.947$ \\
 & &  & &  23.7 & $10.12\pm1.01$ & 29.0 & $ 9.449\pm0.952$ & 37.0 & $ 9.061\pm0.926$ & & \\
 &  &  &  & & & & & & & & \\

2012 Feb 8.2 & 300.2 & 15 & C & 1.3 & $ 1.000\pm0.153$ & 1.8 & $ 1.441\pm0.161$ & 2.5 & $ 1.876\pm0.137$ & 3.5 & $ 2.580\pm0.143$ \\
 & &  & & 5.0 & $ 3.802\pm0.193$ & 7.0 & $ 4.685\pm0.236$ &13.3 & $ 5.596\pm0.561$ &16.0 & $ 5.739\pm0.575$ \\
 & &  & & 23.7 & $ 5.155\pm0.519$ & 29.0 & $ 5.045\pm0.511$ & 37.0 & $ 4.385\pm0.454$ & & \\
 &  &  &  & & & & & & & & \\

2012 Mar 10.1 & 331.1 & 16 & C & 2.5 & $ 1.757\pm0.112$ & 3.5 & $ 2.335\pm0.127$ & 5.0 & $ 3.085\pm0.160$ & 7.0 & $ 3.492\pm0.177$ \\
 & &  & & 13.3 & $ 3.981\pm0.401$ & 16.0 & $ 4.125\pm0.416$ & 23.7 & $ 3.645\pm0.371$ & 29.0 & $ 3.031\pm0.324$ \\
 & &  & & 37.0 & $ 2.826\pm0.329$ & & &  & & & \\
 &  &  &  & & & & & & & & \\

2012 Mar 27.1 & 348.1 & 17 & C & 1.3 & $ 0.529\pm0.176$ & 1.8 & $ 0.927\pm0.125$ & &  & &  \\
 &  &  &  & & & & & & & & \\

2012 Jul 22.7 & 465.7 & 18 & B & 1.4& $ 0.600\pm0.133$ & 1.8& $ 0.592\pm0.098$ & 2.6& $ 0.764\pm0.104$ & 3.5& $ 0.689\pm0.070$ \\
& &  & &  5.0& $ 0.867\pm0.075$ & 7.0& $ 0.812\pm0.074$ & & & & \\
 &  &  &  & & & & & & & & \\

2012 Sep 07.6 & 512.6 & 19 & BnA & 1.4& $ 0.622\pm0.132$ & 1.7& $ 0.562\pm0.077$ & 2.6& $ 0.683\pm0.083$ & 3.5& $ 0.642\pm0.058$ \\
& &  & &  5.0& $ 0.669\pm0.065$ & 7.0& $ 0.496\pm0.048$  & & & & \\
 &  &  &  & & & & & & & & \\

2012 Sep 23.5 & 528.5 & 20 & BnA & 13.3& $ 0.757\pm0.096$ & 17.5& $ 0.682\pm0.104$ & 23.7& $ 0.866\pm0.214$ & 29.0& $ 0.483\pm0.151$ \\
 & &  & & 37.0& $ 0.464\pm0.182$  & & &  & & & \\

\enddata
\tablenotetext{a}{Time since optical discovery of the nova event (2011 April 14; \citealt{Waagan_etal11}).}
\end{deluxetable}

\end{document}